\documentclass[11pt]{article}

\usepackage[utf8]{inputenc}
\usepackage{mathtools}       
\usepackage{graphicx}
\usepackage[colorinlistoftodos]{todonotes}
\usepackage[margin=0.9in]{geometry}
\usepackage{multicol}
\usepackage{graphicx} 
\usepackage{amssymb}
\usepackage{amsmath}
\usepackage{mathrsfs}
\usepackage{physics}
\usepackage{slashed}
\usepackage{tikz-cd}
\usepackage{cite}
\usepackage{hyperref}

\newcommand{\wt}{\widetilde}
\newcommand{\wh}{\widehat}
\newcommand{\NS}{{\rm NS}}

\newcommand{\G}{{\cal G}}
\newcommand{\HH}{{\cal H}}

\newcommand{\XX}{{\cal X}}
\newcommand{\FF}{{\cal F}}

\newcommand{\ben}{\begin{eqnarray}\displaystyle}
	\newcommand{\een}{\end{eqnarray}}
\newcommand{\bea}{\begin{eqnarray}}
	\newcommand{\eea}{\end{eqnarray}}

\newcommand{\be}{\begin{equation}}   
	\newcommand{\ee}{\end{equation}}
  
\newcommand{\refb}[1]{(\ref{#1})}   
  
\def\e{{\epsilon}}

\newcommand{\sectiono}[1]{\section{#1}\setcounter{equation}{0}}

\begin{document}
	\begin{titlepage}
		\rightline{}
		\rightline\today 
		\rightline{MIT-CTP/5875} 
		\begin{center}
			\vskip 1.1cm
			
			{\Large \bf {Gauge algebra and diffeomorphisms in string field theory}}\\

			\vskip 1.5cm
			
			{\large\bf {Raji Ashenafi Mamade and Barton Zwiebach}}
			\vskip 1cm

			\vskip .3cm
			
			{\it   Center for Theoretical Physics -- a Leinweber Institute, 
			\\
				Massachusetts Institute of Technology, \\
				Cambridge MA 02139, USA}\\
			\vskip .1cm
			
			\vskip .4cm
			raji@mit.edu, zwiebach@mit.edu

			\vskip 1.9cm
		\end{center}

		\begin{quote} 	
			
			\centerline{\bf Abstract} 
			\bigskip
			We consider the gauge algebra of closed string field theory with  a focus on diffeomorphisms.  This algebra contains off-shell
information in two ways.  The first way is geometric, through the
choice of three-punctured sphere defining 
the three-string vertex.  We 
establish that to leading order in derivatives the superstring
algebra is universal: identical for any choice of vertex.  
For bosonic strings, however, some off-shell dependence
remains for vertices that require symmetrization.  
Off-shell information also appears because field-dependent redefinition 
of the gauge parameters can alter the algebra.
We analyze this dependence in the language of $L_\infty$ algebras,
looking at the role of trivial gauge transformations 
in the efforts to demonstrate that standard diffeomorphisms are part of
the string gauge symmetry.

			\medskip

		\end{quote} 
		\vfill
		\setcounter{footnote}{0}
		
		\setcounter{tocdepth}{2}  
		
	\end{titlepage}

	\baselineskip 15pt

	\tableofcontents
	
	\sectiono{Introduction and summary}

	In theories of gravity diffeomorphisms  are often represented as coordinate transformations on the 
	 spacetime manifold.  Such gauge
	 transformations are  
	 generated by vector fields.  Tensor fields transform under diffeomorphisms
	  with Lie derivatives and the associated 
	gauge algebra is defined by the Lie bracket of vector fields. 
	
	 The understanding of diffeomorphism symmetry in string theory remains  incomplete. String field theories have an enormous set of gauge symmetries
	 and diffeomorphisms is just one of them.
	In first quantization the appearance of a massless spin two particle in the spectrum and its interactions suggested strongly that string theory would have
	diffeomorphism invariance.  This could be seen explicitly 
	in the
	context of string field theory~(SFT).
		The leading transformation 
		associated to diffeomorphisms of the 
		gravity fluctuation $h_{\mu\nu}$ is given by  
	$\delta h_{\mu\nu} =  \partial_\mu \xi_\nu + 
	\partial_\nu \xi_\mu$, with gauge parameter $\xi_\mu$. Such transformation
	was found to be part of the gauge symmetry
	of free closed string field theory~\cite{Siegel:1985tw}.  
	While it is clear there is an exact symmetry
	in SFT that has a gauge parameter $\xi_\mu$,  the transformations
	one reads from the theory, say on the gravity field, do not take the
	standard form of diffeomorphisms.  Moreover, the gauge algebra has 
	field-dependent terms and only closes on-shell.  
	The question is whether there are redefinitions
	of the gravity field and gauge parameters that allow one to recover
	standard diffeomorphisms.

	This question was sharpened by Ghoshal and Sen~\cite{Ghoshal:1991pu} 
	who described a concrete setup to ascertain if conventional diffeomorphisms
	are consistent with string field theory.  In this setup the effective
	field theory (EFT) of the massless fields, expected to have standard 	
	diffeomorphisms,
	is embedded into the string field theory by finding the string field associated
	to off-shell configurations of the EFT.  Moreover, the gauge parameters of the SFT are expressed in terms of the EFT gauge parameters and field.  One then 
	demands that action of gauge transformations in both theories is consistent with the embedding of the EFT into the SFT.  
	The authors reported no obstruction to carrying this program when including
	cubic (but not higher) terms in the SFT action.  
	
	This identification of diffeomorphisms was done explicitly in the context of 
	double field theory~\cite{Hull:2009mi} including cubic terms of the SFT and working
	out the interactions and gauge transformations to leading order in derivatives. 
	A later analysis, also motivated by double field theory, examined the
	leading $\alpha'$ corrections~\cite{Hohm:2014xsa} to the algebra of
	gauge transformations.  Again, after simplifications by a series of field and parameter redefinitions, the result (eqn.(3.34) of that work), implies a diffeomorphism subalgebra with no corrections to leading order in $\alpha'$.   
	There are no results beyond cubic order in fields. 
	A recent investigation 
	aiming for a manifest description of diffeomorphisms in bosonic
	string theory can be found in~\cite{Mazel:2025fxj}.

The main motivation for our study  here is that 
in the Ramond-Ramond sector
of  type II ~SFT\cite{Sen:2015uaa} diffeomorphisms are realized
in an exotic form, as anticipated in~\cite{Sen:2015nph}.  Another puzzle is
the presence of extra fields that decouple from gravity, allowing nevertheless for
diffeomorphism symmetry.   The derivation
of these exotic transformations from the string field theory
and their analysis is the focus of our
companion paper~\cite{Mamade:2025jbs}.  The preparation for that work included
calculating the diffeomorphism algebra from string field theory, both for bosonic
strings and for type II superstrings.  Such
computations, interesting on their own right, are discussed here.  The algebra
of diffeomorphisms, as we will confirm here, is not exotic (to leading order). 
Nevertheless, in string
field theory it involves off-shell data.  Most of our work here is 
dedicated to understand this off-shell dependence.

The first source of off-shell dependence arises because the 
bracket in the algebra  of gauge transformations is given by 
the string product $[\Lambda_1, \Lambda_2]$
of two string field gauge parameters $\Lambda_1$ and $\Lambda_2$, as in
$[\delta_{\Lambda_2}, \delta_{\Lambda_1} ] = \delta_{[\Lambda_1, \Lambda_2]} + \cdots$. 
This product
is in fact defined by the string theory three-string vertex: a three-punctured sphere with local coordinates around the punctures.  The two gauge parameters go into two of the punctures, and the product is the output at the third puncture.  The gauge 
algebra bracket
manifestly depends on the choice of local coordinates at the punctures, which is
off-shell data in the SFT.   In superstring field theory
there is further off-shell data in the vertex:  the position of a required picture changing operator (PCO).  When speaking about vertices in string field theory there are two possibilities. 
  Ideally, the vertex is symmetric: the local coordinates go into each other (up to phases) by the conformal transformations that permute the punctures.
If the vertex is not symmetric, it is effectively symmetric in the SFT action due to
the symmetrization resulting from the coupling to three identical 
copies of the string field.  Our discussion will
distinguish these two cases.  Note, however, that in the case of the superstring, even
a symmetric vertex is not fully symmetric due to the PCO insertion.
The insertion must be symmetrized explicitly.

The second source of off-shell dependence arises because the algebra of
gauge parameters is affected when one performs field dependent redefinitions
of the gauge parameters.  This, as discussed by~\cite{Ghoshal:1991pu}  and confirmed in
bosonic string theory in~\cite{Hull:2009mi},   adds a new ambiguity.  We discuss this matter in the language of $L_\infty$ algebras, which deals efficiently with 
the non-polynomiality of the string field theory~\cite{Zwiebach:1992ie,Hohm:2017pnh}.  In doing this investigation
we found that a typical redefinition of gauge parameters requires understanding
of trivial gauge symmetries. These are symmetries where the field transformations
are proportional to the equations of motion and thus vanish on-shell.

	\bigskip	
	
The main results in this paper are as follows: 
	
	\begin{enumerate}
	
	\item  The $L_\infty$-based  
		computation of the algebra
		of gauge transformations for type II theories. 
		There are new points, this being a two-field SFT with PCO insertions.  
		This computation, in section~\ref{gaderivx}, 
		uses the framework developed by Firat~\cite{Firat:2024dwt}.
		We describe this gauge algebra in terms of the string products that define the theory. As expected, we find field-dependent structure constants and trivial gauge transformations.  
		
\item  For the bosonic string theory gauge algebra,  
	to leading order in derivatives, the
	bracket of gauge transformations does not involve the off-shell 
	data if the three-string vertex is symmetric. If it is not, and thus
	requires symmetrization, some off-shell dependence remains in the bracket.

	\item  For superstring theory, to leading order in derivatives, the
	bracket of gauge transformations arising in the NSNS sector
	does not involve the off-shell 
	data of the three-string vertex, and is thus {\em universal}. 
	The position of the picture changing operator can be chosen
	arbitrarily but must be symmetrized under the transformations
	that permute the three punctures.  There is no position that avoids
	this symmetrization.   
	
	\item The summary of results for the bracket in both
	bosonic and type II theories is at the end of section~\ref{difeo3ko3if}.
	Ultimately the bosonic string results are more off-shell sensitive because
	the vertex operators associated with the gauge parameters are not primary. In the superstring,
	however, the vertex operators associated to the gauge parameters are primary.
	In both cases, the conformal dimension of the operators is not 
		zero, but this only affects higher derivative contributions. 
				
		\item  
		The algebra
		of diffeomorphisms is essentially 
		standard to leading order in derivatives:  
		after field redefinitions one recovers  
		the familiar Lie bracket of gauge parameters, but one still has field
		dependent structure constants and trivial gauge symmetries.  It remains a challenge to identify a diffeomorphism algebra without such exotic features. 
		
	\item Both the gauge algebra bracket and the trivial gauge transformations
	in the algebra can be modified via field-dependent redefinitions of the gauge parameters and by deforming the standard gauge transformation by the inclusion of trivial transformations. 	
We present such analysis in the framework of $L_\infty$ algebras in section~\ref{gaderivLinf}.  The main
novelty is an analysis of the possible set of trivial gauge transformations in SFT. 
We make progress in classifying such transformations but a complete result
would be desirable. 

\item  We frame the program of Ghoshal and Sen~\cite{Ghoshal:1991pu} in the modern language of
$L_\infty$ algebras and $L_\infty$ morphisms, clarifying the structural 
form of the conditions that establish the identification of 
diffeomorphisms (see section~\ref{froma}).
This may help carry out the program further.

	\end{enumerate}
	
	Basic conventions for calculations are included here; the reader may
	consult section 2 of~\cite{Mamade:2025jbs}  for more details. 
	When not explicitly stated, we follow the conventions of~\cite{Polchinski:1998rr}.

\sectiono{Gauge transformations and $L_\infty$}\label{gautraandLinf} 

The main goal here is to compute the gauge algebra for type II SFT.
This computation is greatly simplified using the analogous 
analysis for bosonic strings~\cite{Zwiebach:1992ie}, as later streamlined in~\cite{Hohm:2017pnh}, both of these done in the language 
of $L_\infty$ algebras.  So we begin by reviewing the key results
for the bosonic string, including the $L_\infty$ `main identity',
gauge transformations, and gauge algebra. (This review also helps set the stage
for the analysis in section~\ref{gaderivLinf}.)  We then turn to
the type II theory, with its own main identity and its novel 
gauge transformations.  In a reformulation
with two-component string fields we recover an $L_\infty$ structure
that then allows us to apply the bosonic string results to the type II
theory.   

\subsection{Bosonic SFT gauge algebra} \label{bossftgaualg}

We use here $L_\infty$ algebras, 
following~\cite{Zwiebach:1992ie,Hohm:2017pnh}.  We have a vector
space graded by a degree that is correlated with Grassmanality,
and graded commutative multilinear products. 
In closed string field theory degree `deg' is related
to ghost number `gh' as deg $= 2- $gh.   
 The degree enters in sign factors where,
for convenience, we omit the `deg' label.  We have, for example:
$ (-1)^{B_1 B_2}  \ \equiv \  (-1)^{{\small\hbox{deg} }(B_1) \cdot {\small\hbox{deg} } (B_2)}$.  Graded commutativity of the products implies that
\be\label{Bcommutation}
[ \cdots B_i , B_j , \cdots ] = (-1)^{B_i B_j}  [ \cdots B_j , B_i , \cdots ]\,.
\ee
  The products, with two or more inputs, 
 satisfy a `main identity'~\cite{Zwiebach:1992ie} that collects the
constraints that define an $L_\infty$ algebra.  With $n\geq 2$ we have,
\be\label{mainide-bos}
		\begin{split}
			& Q[A_1,\cdots,A_n] + \sum_{i=1}^n (-1)^{\e_1+\cdots+\e_{i-1}}[A_1,\cdots,QA_i,\cdots,A_n]  \\   
			& \qquad \, + \sum_{{l\geq 1, k\geq 2}\atop{ l+k =n}}
			\sum_{ {\{i_a; a = 1, \cdots , l\}\atop \{j_b; b = 1, \cdots , k\} } \atop\{i_a\} \cup \{j_b\} = \{1, \cdots, n\}}
			\sigma({i_a},{j_b}) [A_{i_1},\cdots,A_{i_l}, [A_{j_1},\cdots, A_{j_k}]] \ = \ 0 \,,
		\end{split}
		\ee
		where $\sigma({i_a},{j_b})$ is the sign picked up in rearranging $(b_0^- A_1\cdots A_n)$ into $(A_{i_1} \cdots A_{i_l} b_0^- A_{j_1}\cdots A_{j_k})$,
		 and $\e_i$ is the Grassmanality of $A_i$.  The second sum is over inequivalent splittings of $\{1, \cdots, n\}$ to $\{i_1, \cdots, i_l\}$ and  $\{j_1, \cdots, j_k\}$. We do not sum over different permutations of $\{i_a; a = 1, \cdots , l\}$ or different permutations of $\{j_b; b = 1, \cdots , k\} $.  
		 All products have intrinsic degree minus one.
		 
The product with one input is defined by the action of the BRST operator $Q$ operator: $[B] \equiv QB$.  It is useful to include a product $[ \cdot  ]$ with no input, which
for our present applications is set to zero: $[ \, \cdot \, ] \ \equiv \  0$, but in general is just a vector in the state space.
The above main identity, now valid for $n\geq 0$, 
 can be written as 
		\be
		\label{refmainidenc}
		\sum_{{l, k\geq 0}\atop{ l+k =n}} \sum_{ {\{i_a; a = 1, \cdots , l\}\atop \{j_b; b = 1, \cdots , k\} } \atop\{i_a\} \cup \{j_b\} = \{1, \cdots, n\}} \sigma({i_a},{j_b}) [{A}_{i_1},\cdots,{A}_{i_l}, [{A}_{j_1},\cdots, {A}_{j_k}]]= 0\,, \ \ \ \  n \geq 0  \,. 
		\ee
The terms with $l=0$ and those with $k=1$ reproduce
		the terms on the first line of~\refb{mainide-bos}.

For a string field theory expanded around a  
classical solution, we take $[ \, \cdot \, ] \ \equiv \  0$.
The string field $\Psi$ is an element of 
degree zero (Grassmann even). The string field equation 
is ${\cal F}=0$ with  ${\cal F}$ Grassmann odd and given by  
\be
\label{calF-def}
{\cal F} \ = \, \sum_{n=0}^\infty  {1\over n!}\  [\Psi^n] \, = \, 
 \, Q \Psi  + \tfrac{1}{2}[\Psi^2]  + \tfrac{1}{3!} [\Psi^3 ]  + \ldots  \, = \ Q\Psi + \tfrac{1}{2}[ \Psi, \Psi] 
 + \tfrac{1}{3!}[\Psi, \Psi, \Psi] + \ldots \ .
\ee
It is convenient to define primed products that use the string field:
\be\label{primedproducts}
 [  A_1
 \ldots A_n ]' \ \equiv \ \sum_{p=0}^\infty \,  {1\over p!} \, 
 [  A_1
 \ldots A_n \, \Psi^p ]\,,   \qquad  n \geq 1 \,. 
\ee
When there is no risk of confusion,
we omit the commas separating the inputs in the products. 
We write, for example, $
[ A]'  \equiv \ Q' A =  QA  + [A \Psi] +  \ldots$. 
Note also that
\be
[ \, \cdot \, ]' \ \equiv \  {\cal F} \,. 
\ee
The variation of the primed products is rather simple:
\be
\label{var-primed-products}
\delta  [  A_1\ldots A_n ]' \ = \  [  \delta A_1 \ldots A_n ]'
+ \ldots +  [  A_1\ldots \delta A_n ]'
+  [  A_1\ldots  A_n \delta \Psi]' \,.
\ee
A quick calculation then shows that 
\be
\label{varF}
\delta {\cal F} \ = \ Q' (\delta \Psi) \,, 
\ee
To write an action one needs an inner product $\langle \cdot \,, \cdot \rangle$
such that 
\be \label{gcfunct}
\langle \, B_1,  [B_2 , \ldots , B_n] \rangle\, ,   
 \ee
 is a graded commutative function
of all the arguments. This implies, for example,  that 
\be
\langle A , Q B\rangle =  \langle A , [B] \rangle = (-1)^{AB} \langle B , [A] \rangle 
= (-1)^{AB} \langle B, QA \rangle \,.  
\ee
  We also have the basic exchange property  
\be
\label{gr-comm-ip}
\langle  A, B \rangle \ = \ (-1)^{(A+1)(B+1)}  \langle B, A \rangle\,.
\ee
The action is then given by
\be
\label{jnbtflrgs}
S \ = \ \sum_{n=1}^\infty \, {1\over (n+1)!}\langle \Psi \,, \, [\Psi^n]\,  \rangle \,.  
\ee 
We quickly confirm that the variation of the action leads to the field equation 
${\cal F}=0$ for stationarity. 
\be
\label{vary-action}
\delta S  \ = \ \langle \delta \Psi, {\cal F} \rangle \,.
\ee

\medskip
As in~\cite{Hohm:2017pnh} we now derive a family of identities for the primed objects following 
		from the main identity~\refb{refmainidenc}. 
	Taking all the elements to be the Grassmann even $\Psi$, we get
		\be
		\sum_{l\geq 0, k\geq 0 \atop l+k =n}\frac{n!}{l!k!}[\Phi^l [\Phi^k]] = 0, \ \ \  n\geq 0\,. 
		\ee
		Dividing by $n!$ and summing over all $k$ and $l$ we get
		\be
		0 = \, \sum_{l, k\geq 0}\frac{1}{l!k!}[\Psi^l[\Psi^k]] = \sum_{l\geq 0}\frac{1}{l!}[\Psi^l \sum_{k\geq 0} \frac{1}{k} [\Psi^k]] =  \sum_{l\geq 0}\frac{1}{l!}[\Psi^l \mathcal{F}] = Q'\mathcal{F}\,,
		\ee
		implying that 
		\be
		\label{firstprimed}
		Q' {\cal F} = 0 \,. 
		\ee
		Now taking ${A}_1 ={A}$, and ${A}_2 =\cdots = {A}_{n+1} = \Psi $ on the main identity we get
		\be
		\, \sum_{l, k\geq 0\atop l+k =n}\frac{n!}{l!k!} \Bigl(\, (-1)^{{A}}[{A}\Psi^l [\Psi^k]] +  [\Psi^l [{A}\Psi^k]]\, \Bigr)  = 0\,.
		\ee
		Dividing again by $n!$ and summing over all $k,l$ we get
		\be
		0 = \, \sum_{l, k\geq 0}\frac{1}{l!k!} \Bigl(\, (-1)^{{A}}[{A}\Psi^l [\Psi^k]] +  [\Psi^l [{A}\Psi^k]] \Bigr)  =   \sum_{l\geq 0}\frac{1}{l!}\Bigl(\, (-1)^{{A}} [{A}\Psi^l \mathcal{F}] +  [\Psi^l Q'{A}]
		\Bigr) \,. 
		\ee
		With the right hand side written in terms of primed products,
		\be
		(-1)^{{A}}[{A} \mathcal{F}]' +  [Q'{A}]' = 0 \quad \to \quad   
		Q' Q' {\cal A} + [{\cal F A  }]'  = 0 \,. 
		\ee
		Finally, taking ${A}_1$ and ${A}_2$ arbitrary with ${A}_3 = \cdots = {A}_{n+2} = \Psi$,  and repeating the procedure again one finds
		\be\label{main2}
		Q'[{A}_1{A}_2]' + [(Q'{A}_1){A}_2]' +(-1)^{{A}_1}[{A}_1(Q'{A}_2)]' 
		+(-1)^{{A}_1+{A}_2}
		[{A}_1{A}_2\mathcal{F}]' = 0 \,.   
		\ee
The above are the first three identities of the $L_\infty$ algebra
satisfied by the primed products.  They can be in fact read from~\refb{refmainidenc} 
using primes to denote all products and with $[\cdot ] = {\cal F}$.  Including 
an additional identity, we have  
\be
\label{jyndlcsbtt}
\begin{split}
&  0 \ = \ Q' {\cal F}  \,,  \\
&  0 \ = \ Q' ( Q' A) + [{\cal F} A]' \,,  \\
&  0 \ = \ Q' [ A_1 A_2]'  +  [Q'A_1  A_2]'  + (-1)^{A_1} \, [A_1 Q'A_2 ]' + [ {\cal F} A_1 A_2]'  \,, \\
& 0 \ =  \  Q' [A_1  A_2 A_3 ]' \\
& \quad\ +
 [ Q'A_1 A_2 A_3 ]'  
 +  (-1)^{A_1}
[A_1  Q' A_2 A_3  ]'  + (-1)^{A_1+ A_2} [A_1  A_2 Q' A_3 ]'
 \\
 &\quad\ +(-1)^{A_1}    [ A_1  [A_2 A_3]'\,  ]' 
 \ + (-1)^{A_2 (1+A_1)} 
 [ A_2  [A_1 A_3]' \,  ]' \\
 &\quad\ + (-1)^{A_3 (1+A_1+A_2)} 
\,  [ A_3  [\, A_1 A_2]' ]' \ + \  [ {\cal F}  A_1 A_2 A_3]'  \,.
\end{split}
\ee 
The inner product works well with the primed products and this property is inherited from the unprimed products~\refb{gcfunct}. 
One has, for example, the simple property 
\be
\langle Q'A , B\rangle =  (-1)^A \langle A , Q'B \rangle \,.
\ee 
More generally,
\be
\label{stillgradedcomm}
\langle B_1\,, \, [ B_2 \cdots B_n]' \rangle\,, \ \ \hbox{is still graded commutative.}
\ee

\medskip

\noindent
{\bf Standard Gauge transformations:}   
With a Grassmann odd gauge parameter $\Lambda$ the full SFT gauge transformations are simply written as
 \be
\label{vmlvsagdlck}
\delta_\Lambda \Psi \ =  \ Q' \Lambda  . 
\ee
 Using (\ref{varF}) and the second of (\ref{jyndlcsbtt})
\be
\label{jnlvltts}
\delta_\Lambda {\cal F}  \ = \ Q' (\delta_\Lambda \Psi)  \ = \ Q' (Q' \Lambda) 
\ = \  [ \Lambda \,  {\cal F}\,]'  \,.   
\ee
The gauge invariance of the action is immediate.  Since $Q'{\cal F}= 0$ we have
\be
\delta S \ = \  \langle \delta_\Lambda \Psi , {\cal F} \rangle \ = \  
\langle Q' \Lambda , {\cal F} \rangle \ = \ -\langle  \Lambda ,Q'{\cal F} \rangle \ = \ 0 \,.
\ee

\medskip
\noindent
 {\bf Trivial gauge transformations}:   
These are invariances of the action with transformations $\delta \Psi$ 
that that vanish when ${\cal F}=0$.
Two types of 
trivial gauge transformations 
are relevant to the gauge algebra, 
one parameterized by a Grassmann even $\chi$ of ghost number zero and 
another parameterized by two gauge parameters $\Lambda_1, \Lambda_2$.  
They are:
\be
\label{two-kind-of-trivial} 
\begin{split}
\delta_{\chi}^{{}^T} \Psi \ \equiv \ & \   [ \chi\,  {\cal F} ]' \ = \ 
-Q' (Q'\chi) \,, 
\\[1.0ex]
 \delta^{{}^T}_{\Lambda_1, \Lambda_2  } \Psi \ \equiv \ & \   [ \Lambda_1 \Lambda_2 {\cal F} ]' \, . 
\end{split}
\ee
The invariance of the action follows from $\delta S = \langle {\cal F}, \delta \Psi\rangle$
using the graded commutativity of the multilinear function
$\langle B_1 , [ B_2, \cdots B_k] \rangle$ and noting that any product with two ${\cal F}$'s vanishes because ${\cal F}$ is Grassmann odd. 

\medskip 
\noindent
{\bf Gauge algebra:}  The standard gauge transformations form an algebra
that includes the 
trivial gauge transformations
of the second type.  Indeed, assuming the
gauge parameters $\Lambda_1$ and $\Lambda_2$ 
are field independent,  and writing $ \delta_{{\Lambda}_i}\Psi = Q'\Lambda_i = [\Lambda_i]'$, the commutator of gauge transformations is evaluated
as follows 
		\be
[\delta_{\Lambda_2}, \delta_{\Lambda_1}]\Psi =  \delta_{\Lambda_2} [{{\Lambda}}_1]' -  \delta_1 [{{\Lambda}}_2]'  
		= [{{\Lambda}}_1, Q'{{\Lambda}}_2]' -[{{\Lambda}}_2,Q'{{\Lambda}}_1]'\,,
		\ee
		where we used the variation formula~\refb{var-primed-products}.
		 The third equation in~\refb{jyndlcsbtt} enables simplification,
\be
\label{jnexqstvg} 
[\delta_{\Lambda_2}, \delta_{\Lambda_1}]\Psi \  = \  Q'[{{\Lambda}}_1,{{\Lambda}}_2]' + [\, {{\Lambda}}_1,{{\Lambda}}_2, \, {\cal F}]' \,. 
		\ee
Writing the right-hand side in the language of 
gauge transformations, we find the gauge algebra
\be
\label{jnhwlvly}
\bigl[ \delta_{\Lambda_2} \,, \, \delta_{\Lambda_1} \bigr] 
\ = \  \delta_{[\Lambda_1\, \Lambda_2]'}\ + \  \delta^{{}^T}_{\Lambda_1, \Lambda_2 } \,.
\ee
Here $[\Lambda_1, \Lambda_2 ]'$ is the `bracket' that controls the algebra of
gauge transformations.  
 
\subsection{Type II SFT gauge algebra}\label{gaderivx} 
		
		 The analysis here is similar to that in the bosonic string field
		theory~\cite{Zwiebach:1992ie,Hohm:2017pnh}, which we reviewed above. 
		This time, however, there are two variants.  The main
		identity has additional insertions of picture changing operators. Moreover,
		we have two string fields, and thus two gauge parameters.  The two-string-fields situation is dealt with using the notation developed in~\cite{Firat:2024dwt}.  
		
In the type II string field theory, $\HH_{p,q}$  denotes
 the subspace of closed string states with 
	left picture number~$p$ and right picture number~$q$.   The closed
	string fields in this theory are $\Psi$ and $\wt\Psi$ with
		\be
	\label{sschisa}
	\begin{split} 
		& \Psi\in \HH_c \equiv \HH_{-1,-1}\oplus \HH_{-1,-{1\over 2}}\oplus \HH_{-{1\over 2},-1}\oplus \HH_{-{1\over 2},-{1\over 2}}, \\[0.6ex] 
		& \wt\Psi\in \wt\HH_c \equiv
		\HH_{-1,-1}\oplus \HH_{-1,-{3\over 2}}\oplus \HH_{-{3\over 2},-1}\oplus \HH_{-{3\over 2},-{3\over 2}}\, .
	\end{split}
	\ee
	We also have Grassmann even operators  $\XX_0$ and $\bar\XX_0$ 
	that commute with the BRST operator $Q$: 
		\be
	\label{zmpcos}
	\XX_0=\ointop {dz\over z} \XX(z), \qquad \bar\XX_0=\ointop {d\bar z\over \bar z} \bar\XX(\bar z)\, , 
	\ee
		where $\XX(z)$ and $\bar\XX (\bar z)$ are picture changing operators.
Finally, we have an operator ${\cal G}: \wt\HH_c \to \HH_c$ whose definition is given by the action on the various subspaces of $\wt\HH_c$ as follows:
\be
	\label{Gdefpcc}
	\G\equiv \begin{cases} 
	\hbox{${\bf 1}\ $   on $\HH_{-1,-1}$,}\cr
	\hbox{$\XX_0$ on $\HH_{-1,-3/2}$,} \cr 
	\hbox{$\bar\XX_0$ on $\HH_{-3/2,-1}$,}\cr  
	\hbox{$\bar\XX_0\XX_0$ on $\HH_{-3/2,-3/2}$.} 
	\end{cases}
	\ee 
	With Grassmann odd gauge parameters $(\Lambda, \wt\Lambda) \in (\HH_c, \wt\HH_c)$ at ghost number one, the gauge  transformations take the form
	\be
	\label{gaugetrsft}
	\begin{split}
	\delta_{\Lambda, \wt\Lambda}  |\Psi\rangle  =& \  Q | \Lambda\rangle  + \sum_{n=1}^\infty  {1\over n!}  \, {\cal G} \,  [ \Lambda \Psi^n ]\,,    \\  
				\delta_{\Lambda, \wt\Lambda}  |\widetilde\Psi\rangle  =& \  Q |\widetilde \Lambda\rangle  
+ \sum_{n=1}^\infty  {1\over n!}  \, [ \Lambda \Psi^n ]\,.  \\  
	\end{split}
	\ee
The action is given by 
\be\label{ebvii}
	S=- \tfrac{1}{2}  \langle \wt \Psi ,  
	Q \, \G \, \wt\Psi\rangle  
	+ \langle \wt \Psi ,   
	Q \, \Psi\rangle +\sum_{n=3}^\infty {1\over n!} \langle \Psi, [\Psi^{n-1}] \rangle \, .
	\ee

The main identity for the superstring field theory is 
given in~\cite{Sen:2015uaa}, eqn.~(2.15):
		\be\label{mainide}
		\begin{split}
			& Q[A_1,\cdots,A_n] + \sum_{i=1}^n (-1)^{\e_1+\cdots+\e_{i-1}}[A_1,\cdots,QA_i,\cdots,A_n]  \\   
			& \qquad \, + \sum_{{l\geq 1, k\geq 2}\atop{ l+k =n}}
			\sum_{ {\{i_a; a = 1, \cdots , l\}\atop \{j_b; b = 1, \cdots , k\} } \atop\{i_a\} \cup \{j_b\} = \{1, \cdots, n\}}
			\sigma({i_a},{j_b}) [A_{i_1},\cdots,A_{i_l}, \mathcal{G}[A_{j_1},\cdots, A_{j_k}]] \ = \ 0 \,.
		\end{split}
		\ee
		This is almost the same as~\refb{mainide-bos}: all sign factors and 
		splittings are defined in identical fashion.   The only difference is
		the appearance of the $\G$ operator acting on the nested product.

As in the case of bosonic strings, we are only 
concerned with the classical theory and thus  
all products arise from genus zero surfaces. 
		We also define the product with one input as the BRST
		operator and set the product with no input equal to zero:
		\be    
		[A] \equiv  QA  \,, \ \ \ \  [\, \cdot\, ] = 0 \,.
		\ee

		Our first goal is to get for this theory the analog of the simpler form~\refb{refmainidenc} of the main identity. 
		Since the state space of the SFT includes states in $\HH_c$ and states in $\wt \HH_c$, we  define string products acting on two-component elements ${\cal A}$, denoted with calligraphic symbols.  We write
		\be
		\label{qandzeroact}
		\mathcal{A} = \begin{pmatrix}
			A\\ \tilde A
		\end{pmatrix} \ \ \ \hbox{with} \  \   A\in \HH_c, \ \ \wt A \in \wt\HH_c \,, \ \ \ 
		\e_A = \e_{\wt A}  \equiv \e_{\cal A} \,,  
		\ee
		requiring both components of the element ${\cal A}$ to have the same Grassmanality.
		We now have the lowest products
		\be
		[{\cal A} ] \equiv  Q {\cal A}  = \begin{pmatrix}
			QA\\[0.5ex] Q\tilde A 
		\end{pmatrix}\,, \ \ \ \ [\, \cdot\,] \equiv \begin{pmatrix}
			0\\[0.5ex] 0 
		\end{pmatrix} = 0 \,. 
		\ee
		For products with two or more inputs  ${\cal A}_i =  \begin{pmatrix}
			A_i\\ \wt A_i 
		\end{pmatrix}$,  we declare
		\be
		\label{deftwo-brackets} 
		[\mathcal{A}_1,\cdots,  \mathcal{A}_n] \equiv \begin{pmatrix}
			\mathcal{G} [A_1, \cdots A_n] \\[0.4ex] [A_1, \cdots A_n]  \end{pmatrix} \,, \ \ \ 
		n\geq 2 \,.
		\ee
		Note that the $\wt A_i$ do not appear in the $n\geq 2$ products, they only show up under
		the $Q$ action, as in~\refb{qandzeroact}. 
		The main identity~\refb{mainide} can be written as 
		\be
		\label{refmainiden}
		\sum_{{l, k\geq 0}\atop{ l+k =n}} \sum_{ {\{i_a; a = 1, \cdots , l\}\atop \{j_b; b = 1, \cdots , k\} } \atop\{i_a\} \cup \{j_b\} = \{1, \cdots, n\}} \sigma({i_a},{j_b}) [\mathcal{A}_{i_1},\cdots,\mathcal{A}_{i_l}, [\mathcal{A}_{j_1},\cdots, \mathcal{A}_{j_k}]]= 0 \,. 
		\ee
		In here, the main identity~\refb{mainide} appears in the $\wt\HH_c$ part of the identity. The $\HH_c$ part of the identity is the main identity~\refb{mainide} multiplied by ${\cal G}$, and it forms an algebraic structure called a twisted $L_\infty$ algebra~\cite{Singh:2024mek}.  Note that the terms with $l=0$ and those with $k=1$ reproduce the terms on the first line of~\refb{mainide}.  This is the rewriting we wanted.

In this two-component formalism the string field, denoted by $\Phi$,
is  Grassmann even and incorporates $\Psi$ and $\wt\Psi$ as follows 
\be
\Phi = \begin{pmatrix}
			\Psi \\ \tilde \Psi
		\end{pmatrix} \,.
		\ee
As for bosonic strings we make use of primed products, 	
defined in a completely analogous way: 
		\be
		[\mathcal{A}_1,\cdots,\mathcal{A}_n]' \equiv \sum_{p=0}^{\infty}\tfrac{1}{p!}[\mathcal{A}_1,\cdots,\mathcal{A}_n,\Phi^p]\,, \ \ \  n \geq 0\,. 
		\ee
Under variations $\delta {\cal A}_i$ and $\delta \Phi$ the primed products change
as they did for bosonic strings in~\refb{var-primed-products} 
		\be
		\label{varprimedprods}
		\delta [\mathcal{A}_1,\cdots,\mathcal{A}_n]' = 
		\sum_{i=1}^n [\mathcal{A}_1,\cdots,\delta \mathcal{A}_i, 
		\cdots, \mathcal{A}_n]'  
		+ [\mathcal{A}_1,\cdots,\mathcal{A}_n,\delta \Phi]'\,.
		\ee
It is useful to introduce an inner product on the  space of
two-component string fields. We define
\be
\langle {\cal A} , {\cal B} \rangle \equiv \,  - \langle \wt A, {\cal G} \wt B \rangle 
+ \langle \wt A , B \rangle  +   \langle A , \wt B\rangle \,, \ \ \hbox{with} \ \ 
{\cal A} =\begin{pmatrix}
			A\\[0.5ex] \wt A 
		\end{pmatrix}\,,  \ \ {\cal B} =\begin{pmatrix}
			B\\[0.5ex] \wt B     
		\end{pmatrix}   \,. 
\ee
We use the same symbol for the inner product in both the original and the two-component space, the distinction can be made by looking at the inputs. 
The action in the two component notation is now easy to write.  Using the new inner product one quickly verifies that the action~\refb{ebvii}
becomes 
\be\label{ebviixx}
	S= \tfrac{1}{2}  \langle \Phi ,  
	Q \Phi\rangle  
	+\sum_{n=3}^\infty {1\over n!} \langle \Phi , [\Phi^{n-1}] \rangle \, .
	\ee
This is of exactly the same form as the action in 
 bosonic closed string field theory.

\medskip		
For the primed product with no input we also use the notation developed for bosonic strings.  This time the object ${\cal F}$ is a two-component set of field equations: 
\be
\label{Ftwo-comp}
 \mathcal{F} = \begin{pmatrix}
					\mathcal{E} \\  \wt {\mathcal{E}}
				\end{pmatrix}\equiv [\cdot ]' = 
\sum_{n=1}^{\infty}\tfrac{1}{n!}[\Phi^n] =  
\begin{pmatrix} Q\Psi +\sum_{n=2}^\infty{1\over n!} \mathcal{G}[\Psi^n] 
 \\[1.7ex] 
 Q\tilde \Psi + \sum_{n=2}^\infty{1\over n!}[\Psi^n] \end{pmatrix} \,.
\ee
We can relate
${\cal E}$ and $\wt{\cal E}$  to
the field equations arising from direct variation of the 
action~\refb{ebvii}.  A little work gives
\be
\label{vareomII}
\delta S =  \langle \delta \wt \Psi, E_{\wt \Psi} \rangle  +  
 \langle \delta \Psi, E_{\Psi} \rangle\,, 
 \ee
 with
 \be
 E_{\wt \Psi } =   Q \Psi - Q {\cal G} \wt \Psi\,, \ \ \ \ 
 E_\Psi =   Q \wt \Psi + \sum_{n=2}^\infty {1\over n!} [ \Psi^n] \,. 
 \ee
Comparing with the expressions for ${\cal E}$ and $\wt{\cal E}$ in~\refb{Ftwo-comp}, 
 we note that
 \be
 E_\Psi  = \wt{\cal E}   \   , \ \ \    \ \ E_{\wt \Psi} =  {\cal E} -  {\cal G} \wt{\cal E}\,.  
 \ee
 Note that ${\cal E} =  E_{\wt\Psi} + {\cal G} \wt E_\Psi$ is the linear combination of equations of motion from which $\wt\Psi$ drops out;  it is the equation of motion
 of the dynamical, interacting string field.  
 Back in the action variation~\refb{vareomII} we have
 \be
 \label{sdeltaFII} 
 \delta S \ = \   \langle \delta \wt \Psi, \,  {\cal E} -  {\cal G} \wt {\cal E}  \rangle  +  
 \langle \delta \Psi,  \wt {\cal E} \rangle  \ =\  \langle \delta \Phi , {\cal F} \rangle  \,,
 \ee
 using the two-component inner product.  The relation
 $\delta S =  \langle \delta \Phi , {\cal F} \rangle$ is essential for a simple understanding of the gauge invariance of the action and the treatment of trivial gauge transformations.

Since the type II main identity~\refb{refmainiden} is formally identical to the bosonic
string main identity~\refb{refmainidenc} and we use exactly the same definition
for primed products, $Q'$ and ${\cal F}$ (with the bosonic $\Psi$ replaced
by the type II  string field $\Phi$), the primed products here satisfy the same identities
as the bosonic string primed products.

		\medskip   
It is clear from the algebraic structures that $\delta \Phi = Q' \, \wh \Lambda$, for 
a two-component gauge parameter $\wh\Lambda$ is a gauge symmetry leaving
the action invariant.  Writing:
\be
\wh\Lambda = \begin{pmatrix}
			\Lambda \\[0.3ex] \wt\Lambda
		\end{pmatrix}  \,,
\ee
we have
		\be
		\delta_{\wh\Lambda}\Phi \, = \,   
		 \begin{pmatrix} \delta_{\Lambda, \wt\Lambda }\Psi 
			\\[1.0ex]
			\delta_{\Lambda, \wt\Lambda} \wt \Psi \end{pmatrix} 
		\ = Q' \wh\Lambda  = [\wh\Lambda]' = \begin{pmatrix}Q\Lambda +   \sum_{n=1}^{\infty}\tfrac{1}{n!} \mathcal{G}[\Lambda,\Psi^n] \\[1.8ex]   Q \wt \Lambda +   \sum_{n=1}^{\infty}\tfrac{1}{n!}[\Lambda,\Psi^n]\end{pmatrix} \,.
		\ee
		Not surprisingly, these are the 
		anticipated gauge transformations in~\refb{gaugetrsft}. 
		
		\medskip
		Since the gauge transformations take the same form as those of bosonic string theory (both use $Q'$) and the primed identities are the same for the two 
		theories, the algebra results for the bosonic string hold for type II, when using the two-component string fields and gauge parameters.  Thus, equation~\refb{jnexqstvg} becomes, with  $\delta_i \equiv \delta_{\wh{\Lambda}_i}$,	\be
		\begin{split}
			[\delta_2, \delta_1]\Phi &\  = \  Q'[\, {\wh{\Lambda}}_1,{\wh{\Lambda}}_2]' + [\, {\wh{\Lambda}}_1,{\wh{\Lambda}}_2, \, {\cal F}]' \,. 
		\end{split}
		\ee
		The first term on the right-hand side is a gauge transformation with gauge parameter $\hat\Lambda_{12}$:
		\be
		\wh\Lambda_{12} \  =\ \begin{pmatrix}
			\Lambda_{12} \\[0.7ex]  \wt \Lambda_{12}
		\end{pmatrix} \ = \ [{\wh{\Lambda}}_1,{\wh{\Lambda}}_2]' = \begin{pmatrix}
			\mathcal{G} [\Lambda_1, \Lambda_2]' \\[0.7ex]   [\Lambda_1, \Lambda_2]' 
		\end{pmatrix}\,.
		\ee
		Writing $\delta_{\hat \Lambda} =  \delta_{\Lambda, \wt\Lambda}$, we have, explicitly,
		\be
		\begin{split}
			& [\delta_{\Lambda_2, \widetilde \Lambda_2} \, , 
			\delta_{\Lambda_1, \widetilde \Lambda_1}]\, \Psi 
			=\  \delta_{\Lambda_{12}, \widetilde \Lambda_{12}} \Psi + \mathcal{G}[\Lambda_1,\Lambda_2,  \mathcal{E}]'\,, \\[1.0ex]
			& [\delta_{\Lambda_2, \widetilde \Lambda_2} \, , 
			\delta_{\Lambda_1, \widetilde \Lambda_1}]\, \wt\Psi = \ 
			\delta_{\Lambda_{12}, \widetilde \Lambda_{12}} \wt\Psi +[\Lambda_1,\Lambda_2,\mathcal{E}]'\,. 
		\end{split}
		\ee
		As indicated by the primes, the gauge parameters $\Lambda_{12}$  and 
		$\wt\Lambda_{12}$ in the gauge algebra are field dependent. 
		Note that the original tilde gauge parameters 
		$\wt \Lambda_1$ and $\wt \Lambda_2$ in $\delta_1$ and $\delta_2$
		do not appear in the expressions
		for $\Lambda_{12}$ and $\wt \Lambda_{12}$. 
		
		The structure of trivial transformations is completely analogous to that of bosonic string theory because the variation of the action is $\langle \delta \Phi, {\cal F}\rangle$,  ${\cal F}$ is odd, and all identities are of the same form.  Due to~\refb{deftwo-brackets} the tilde components of the gauge parameters do not appear in trivial 
		transformations, and nor does
		$\wt {\cal E}$.

\sectiono{Diffeormorphism algebra from SFT}

We have seen that the commutator of two gauge transformations
results in a gauge transformation with a field-dependent gauge parameter,
along with terms that vanish when the equation of motion holds. 
In the case of bosonic SFT, we have
\be\label{bosalg}
[\delta_{\Lambda_2}, \delta_{\Lambda_1}] \Psi
= \delta_{\Lambda_{12}} \Psi + [\Lambda_1, \Lambda_2, \mathcal{F}]',
\ee
and in the case of type II SFT,
\be\label{superalg}
[\delta_{\widehat \Lambda_2}, \delta_{\widehat \Lambda_1}] \Phi 
= \delta_{\widehat \Lambda_{12}} \Phi + [\widehat{\Lambda}_1, \widehat{\Lambda}_2, \mathcal{F}]'.
\ee
In this section, we compute the field-independent part of the gauge parameter $\Lambda_{12}$ and $\widehat \Lambda_{12}$
in the massless sectors of both bosonic and superstring field theory. 
We will use the following normalizations for the bosonic and superstring cases, respectively:
\be\label{normalizationb}
\begin{split} 
	\left\langle c\bar{c}(z_1)\, c\bar{c}(z_2)\, c\bar{c}(z_3)\, e^{ip\cdot X} \right\rangle 
	= - |z_{12} z_{13} z_{23}|^2 (2\pi)^D \delta^D(p),
\end{split}
\ee
\be\label{normalizations}
\begin{split} 
	\left\langle c\bar{c}(z_1)\, c\bar{c}(z_2)\, c\bar{c}(z_3)\, e^{-2\phi} e^{-2\bar{\phi}} e^{ip\cdot X} \right\rangle 
	= - |z_{12} z_{13} z_{23}|^2 (2\pi)^D \delta^D(p).
\end{split}
\ee
We will also make use of the picture changing operator, whose holomorphic part 
${\cal X}(z)$ has picture number one, ghost number zero, and conformal dimension zero:
\be
\mathcal{X}(z) \equiv \{ Q, \xi(z) \} = c \partial \xi + e^{\phi} T_F 
- \partial \eta\, e^{2\phi} b - \partial(\eta\, e^{2\phi} b), \hskip25pt 
\mathcal{X}_0 \equiv \oint \frac{dz}{z} \mathcal{X}(z).
\ee
The operators $\bar{\cal X}(\bar z)$ and $\bar{\cal X}_0$ are similarly defined.

\subsection{Diffeomorphism algebra in the bosonic string}
We now consider the algebra of gauge transformations in the bosonic theory. The gauge parameter in bosonic string field theory is given by
\be
\Lambda = \int \frac{d^D p}{(2\pi)^D} \Bigl( 
\lambda_{\mu}(p)\, c\, \partial X^\mu    
- \bar{\lambda}_{\mu}(p)\, 
\bar{c}\, \bar{\partial} X^\mu 
+ \mu(p)\, (\partial c + \bar{\partial} c) \Bigr) e^{ip \cdot X} \,.
\ee
The gauge parameters here are $\lambda_\mu, \bar\lambda_\mu$, and $\mu$. 
The diffeomorphism parameter $X^\mu$  and the Kalb-Ramond gauge parameter
$\epsilon_\mu$ are known to be given by
\be
X_\mu \equiv \tfrac{1}{2}(\lambda_\mu + \bar \lambda_\mu ) \,, \ \ \ 
\e_\mu\equiv \tfrac{1}{2}(\lambda_\mu -\bar \lambda_\mu )\,.
\ee

We now build operators dual in the inner product  
to the states in $\Lambda$. 
These are three operators, $\Phi_\mu(p)$, $\bar \Phi _\mu(p)$,
and $\Phi (p)$ satisfying: 
\be
\label{dualbos} 
\expval{\Phi_\mu(p), \Lambda}  =  \lambda_\mu(-p)\, ,
\ \ \ 
\expval{\bar {\Phi}_\mu(p), \Lambda}  = \bar  \lambda_\mu(-p)\, , 
\ \ \ 
\expval{\Phi(p), \Lambda}  =  \mu(-p)\, . 
\ee
It is straightforward to check that the operators are:
\be
\begin{split}
\Phi^\mu(p) & = {\bar \partial}^2\bar c (\partial c + \bar \partial \bar c)c \bar c \partial X^\mu  e^{ip\cdot X}\,  ,\\[1.0ex]
	\bar \Phi^\mu(p) & = \partial^2 c (\partial c + \bar \partial \bar c)c \bar c \bar \partial X^\mu  e^{ip\cdot X}\, ,\\[1.0ex]
\Phi (p)\  &  =  \partial^2 c {\bar \partial}^2\bar c c\bar c\, e^{ip\cdot X} \,.
\end{split}
\ee
The gauge parameter on the right-hand side of the algebra~\refb{bosalg} is given by
\be\label{bosbos}
\Lambda_{12} = [\Lambda_1, \Lambda_2]' = [\Lambda_1, \Lambda_2] + {\cal O}(\Psi)\,,
\ee
where the primed products are defined as in~\refb{primedproducts}, with $\Psi$ denoting the bosonic string field. We will focus here on the field-independent part, $[\Lambda_1, \Lambda_2]$, which we denote by $\Lambda_{12}$ with a slight abuse of notation. Since we are focusing on diffeomorphisms, we compute the product $[\Lambda_1, \Lambda_2]$ under the assumption $\mu_1(p) = \mu_2(p) = 0$. We denote the component fields of $[\Lambda_1, \Lambda_2]$ by $\lambda_{12}^\mu(p)$ and $\bar{\lambda}_{12}^\mu(p)$.

Since we are doing field theory, our computations are off-shell. 
The CFT operators
are inserted using local coordinates, and we need to do conformal transformations.  In the three-string multilinear function the operators are inserted at $w_i = 0$,
the origin in the local coordinates:
\be
\begin{split}
	\{  O_1, O_2, O_3\} = & \  \langle  \ O_1 (w_1=0) \  O_2 (w_2 = 0 ) \ O_3 (w_3=0) \rangle 
	\\[1.0ex]
	= &  \  \langle  f_1 \circ  O_1 (z_1)  f_2 \circ O_2 (z_2 )  f_3 \circ O_3 (z_3 ) \rangle\, .
\end{split}
\ee 
 We distinguish between $\lambda^{12}_\mu(-p) = \{ \Phi^\mu(p), \Lambda_1, \Lambda_2 \}$ defined with respect 
         to the symmetric string multilinear function computed using symmetric string vertices and $\lambda_{12}^{'\mu}(-p) = \{ \Phi^\mu(p), \Lambda_1, \Lambda_2 \}'$  
 computed with string vertices with arbitrary local coordinates.  We start by computing the latter, which reduces to the former when evaluated on a symmetric vertex, or when the result is explicitly symmetrized over permutations of the punctures.
\be
\begin{split}
	\lambda_{12}^{'\mu}(-p) =  \ \{  \Phi^\mu(p), \Lambda_1, \Lambda_2\}'   = & \  \langle  \  \Phi^\mu(p) (w_1=0) \  \Lambda_1 (w_2 = 0 ) \ \Lambda_2 (w_3=0) \rangle 
	\\[1.0ex]
	= &  \  \langle  f_1 \circ   \Phi^\mu(p) (z_1)  f_2 \circ \Lambda_1 (z_2 )  f_3 \circ \Lambda_2 (z_3 ) \rangle\, .
\end{split}
\ee
We then have
\be\label{bossal}
\lambda_{12}^{'\mu}(-p) =  
\int {d^D p_2\over (2\pi)^D}{d^D p_3\over (2\pi)^D}  
\Bigl(
\lambda_{1\alpha}(p_2)\, \bar{\lambda}_{2\beta}(p_3)\, Z_1^{\mu\alpha\beta}
+ \bar{\lambda}_{1\alpha}(p_2)\, \lambda_{2\beta}(p_3)\, Z_2^{\mu\alpha\beta}
+ \lambda_{1\alpha}(p_2)\, \lambda_{2\beta}(p_3)\, Z_3^{\mu\alpha\beta}
\Bigr),
\ee
where $Z_i^{\mu\alpha\beta} = Z_i^{\mu\alpha\beta}(p, p_2, p_3)$ for $i = 1, 2, 3$, and
\be \label{ZZeqns}
\begin{split}
	Z_1^{\mu\alpha\beta} &= \left\langle 
	f_1 \circ \bar{\partial}^2 \bar{c}\, \partial c\, c \bar{c}\, \partial X^\mu\, e^{i p \cdot X}(z_1) \,\,
	f_2 \circ c \partial X^\alpha\, e^{i p_2 \cdot X}(z_2) \,\,
	f_3 \circ \bar{c} \bar{\partial} X^\beta\, e^{i p_3 \cdot X}(z_3)
	\right\rangle, \\[1.0ex]
	Z_2^{\mu\alpha\beta} &= \left\langle 
	f_1 \circ \bar{\partial}^2 \bar{c}\, \partial c\, c \bar{c}\, \partial X^\mu\, e^{i p \cdot X}(z_1) \,\,
	f_2 \circ \bar c \bar \partial X^\alpha\, e^{i p_2 \cdot X}(z_2) \,\,
	f_3 \circ c \partial X^\beta\, e^{i p_3 \cdot X}(z_3) 
	\right\rangle, \\[1.0ex]
	Z_3^{\mu\alpha\beta} &= \left\langle 
	f_1 \circ \bar{\partial}^2 \bar{c}\, \bar{\partial} \bar{c}\, c \bar{c}\, \partial X^\mu\, e^{i p \cdot X}(z_1) \,\,
	f_2 \circ c \partial X^\alpha\, e^{i p_2 \cdot X}(z_2) \,\,
	f_3 \circ c \partial X^\beta\, e^{i p_3 \cdot X}(z_3) 
	\right\rangle.
\end{split}
\ee
All other terms vanish by ghost number conservation.

Using the conformal transformations from~\refb{conftranf} and separating the ghost and matter parts, the $Z_1$ correlator is written as:
\be
\begin{split}
	Z_1^{\mu\alpha\beta}
	= & \left\langle 
	\left(\bar{\partial}^2 \bar{c}\, \partial c\, c\, \bar{c} 
	+ \tfrac{\bar f_1''}{\bar f_1'^2} \partial c\, \bar{\partial} \bar{c}\, c\, \bar{c} \right)(z_1)\,  
	c(z_2)\, \bar{c}(z_3)
	\right\rangle M^{\mu\alpha\beta}\,,
\end{split}
\ee
where 
\be
M^{\mu\alpha\beta}  =  \Bigl\langle 
\Bigl( \partial X^\mu - \tfrac{ip^\mu}{4} \tfrac{f_1''}{f_1'^2} \Bigr) 
e^{ip\cdot X}(z_1)\,
\Bigl( \partial X^\alpha - \tfrac{ip_2^\alpha}{4} \tfrac{f_2''}{f_2'^2} \Bigr)
e^{ip_2\cdot X}(z_2)\,
\Bigl( \bar{\partial} X^\beta - \tfrac{ip_3^\beta}{4} \tfrac{\bar{f}_3''}{\bar{f}_3'^2} \Bigr)
e^{ip_3\cdot X}(z_3)
\Bigr\rangle\,.
\ee
We dropped the conformal factors which only contribute to the second order in momentum. The ghost correlator is readily computed and we thus have:
\be 
	Z_1^{\mu\alpha\beta} = 2 \bar{z}_{13} z_{12}^2 \left( 1 - \bar{z}_{13} \tfrac{\bar{f}_1''}{2 \bar{f}_1'^2} \right) M^{\mu\alpha\beta}\,.
\ee
We now compute the matter correlator to leading order in derivatives, that is, to first order. In this correlator, the two $\partial X$ fields must be contracted. If each $\partial X$ contracts with an exponential, the result is at least second order in momenta. Therefore, to obtain the leading term, $\partial X^\mu$ and $\partial X^\alpha$ must contract with each other, while $\bar{\partial} X^\beta$ contracts with one of the exponentials. The off-shell parts of the first and second operators contain explicit momenta and thus do not contribute to leading order. Thus, to leading order:
\be
\begin{split}
	M^{\mu\alpha\beta}
	&= -\frac{\eta^{\mu\alpha}}{2 z_{12}^2}
	\Bigl( \frac{ip^\beta}{2 \bar{z}_{13}} 
	+ \frac{ip_2^\beta}{2 \bar{z}_{23}} 
	- \frac{ip_3^\beta}{4} \tfrac{\bar{f}_3''}{\bar{f}_3'^2} \Bigr) 
	(2\pi)^D \delta\left( \sum p \right)
	+ \order{p^2} \,, \\ 
	&= \frac{\eta^{\mu\alpha}}{4 z_{12}^2}
	\Bigl( -\tfrac{\bar{z}_{12}}{\bar{z}_{13} \bar{z}_{23}}\, ip_2^\beta 
	+ \bigl( \tfrac{1}{\bar{z}_{13}} + \tfrac{\bar{f}_3''}{2 \bar{f}_3'^2} \bigr) ip_3^\beta \Bigr)
	(2\pi)^D \delta\left( \sum p \right)
	+ \order{p^2}\,.
\end{split}
\ee
Combining the ghost and matter parts, we obtain:
\be
\begin{split}
	Z_1^{\mu\alpha\beta} 
	&= \tfrac{1}{2} \eta^{\mu\alpha} 
	\Bigl( \bar{A}_{2,13}\ i p_2^\beta 
	+ \bar{A}_{2,31} \bar{A}_{2,13}\ i p_3^\beta \Bigr)
	(2\pi)^D \delta\left( \sum p \right) + \order{p^2}\,,
\end{split}
\ee
where we defined the combination
\be\label{aijk}
A_{i,jk} \equiv \frac{z_{ij}}{z_{ik}} - \frac{f_j''}{2 f_j'^2} \frac{z_{ij} z_{jk}}{z_{ik}}
= \frac{z_{ij}}{z_{ik}} \Bigl( 1 - \frac{f_j''}{2 f_j'^2} z_{jk} \Bigr)\,,
\ee
and the corresponding complex conjugate $\bar A_{i,jk}$.
Note that in all instances of our calculations $i \not= j \not= k$; their values are a permutation
of $\{ 1,2,3\}$.

The second correlator in~\refb{ZZeqns} is the negative of the first, with the substitutions \( z_2 \leftrightarrow z_3 \), \( p_2 \leftrightarrow p_3 \), and \( \alpha \leftrightarrow \beta \). Therefore, we find:
\be
Z_2^{\mu\alpha\beta} 
= -\tfrac{1}{2} \eta^{\mu\beta} 
\left( \bar{A}_{3,12}\  i p_3^\alpha 
+ \bar{A}_{3,21} \bar{A}_{3,12}\  i p_2^\alpha \right)
(2\pi)^D \delta\left( \sum p \right) + \order{p^2}\,.
\ee

Factorizing the ghost and matter parts, and using the conformal transformations from~\refb{conftranf}, the third correlator is written as:
\be
\begin{split}
	Z_3^{\mu\alpha\beta} &= \Bigl\langle {\bar \partial}^2 \bar c\, \bar \partial \bar c\, c \bar c(z_1) c(z_2) c(z_3) \Bigr\rangle N^{\mu\alpha\beta} = 2 z_{12} z_{13} z_{23} \ N^{\mu\alpha\beta} \, ,
\end{split}
\ee
where
\be
N^{\mu\alpha\beta} = \Bigl\langle \Bigl(\partial X^\mu - \tfrac{ip^\mu}{4} \tfrac{f_1''}{f_1'^2} \Bigr) e^{ip\cdot X}(z_1)
\Bigl(\partial X^\alpha - \tfrac{ip_2^\alpha}{4} \tfrac{f_2''}{f_2'^2} \Bigr) e^{ip_2\cdot X}(z_2)
\Bigl(\partial X^\beta - \tfrac{ip_3^\beta}{4} \tfrac{f_3''}{f_3'^2} \Bigr) e^{ip_3\cdot X}(z_3) \Bigr\rangle\, . 
\ee
For this matter correlator, the leading first-order contribution in momentum arises when two of the \( \partial X \) fields contract with each other, and the third contracts with a momentum exponential. The result is then
\be
\begin{split}
	Z_3^{\mu\alpha\beta} & = 2 z_{12} z_{13} z_{23} \Bigl[ -\frac{\eta^{\mu\alpha}}{2 z_{12}^2}
	\Bigl( \tfrac{ip^\beta}{2 z_{13}} + \tfrac{ip_2^\beta}{2 z_{23}} - \tfrac{ip_3^\beta}{4} \tfrac{f_3''}{f_3'^2} \Bigr)
	- \frac{\eta^{\mu\beta}}{2 z_{13}^2}
	\Bigl( \tfrac{ip^\alpha}{2 z_{12}} - \tfrac{ip_3^\alpha}{2 z_{23}} - \tfrac{ip_2^\alpha}{4} \tfrac{f_2''}{f_2'^2} \Bigr) \\
	& \hskip65pt - \frac{\eta^{\alpha\beta}}{2 z_{23}^2}
	\Bigl( -\tfrac{ip_3^\beta}{2 z_{13}} - \tfrac{ip_2^\beta}{2 z_{12}} - \tfrac{ip^\beta}{4} \tfrac{f_1''}{f_1'^2} \Bigr) \Bigr] (2\pi)^D \delta( \sum p ) + \order{p^2}\,,  \\
	&= \tfrac{1}{2} \Big[ \eta^{\mu\alpha} \Bigl( -ip_2^\beta - A_{2,31}\ ip_3^\beta \Bigr)
	+ \eta^{\mu\beta} \Bigl( ip_3^\alpha + A_{3,21}\ ip_2^\alpha \Bigr) \\
	&\hskip20pt	+ \eta^{\alpha\beta} \Bigl( A_{3,12}\ ip_2^\mu - A_{2,13}\ ip_3^\mu \Bigr) \Big] (2\pi)^D \delta( \sum  p ) + \order{p^2}\,.
\end{split}
\ee
Collecting the results for the $Z_i$'s, and writing equation~\refb{bossal} in position space we now have
\be\label{bprimed}
\begin{split}
	\lambda_{12}^{'\mu}	& = \tfrac{1}{2}\, \bigl[ \  \bar A_{3,12} \, 
	\bar \lambda_1\cdot\partial\lambda_2^\mu 
	-\bar A_{2, 13}\, 
	\bar \lambda_2\cdot\partial\lambda_1^\mu \ \ - \ \
	\ \bar A_{2,31} \bar A_{2,13} \, 
	\partial \cdot \bar \lambda_2 \lambda_1^\mu 
	\ + \bar A_{3,21} \bar A_{3,12} \, 
	\partial \cdot \bar \lambda_1 \lambda_2^\mu \\[1.0ex]
	& \qquad  	+   \lambda_1\cdot\partial\lambda_2^\mu - \lambda_2\cdot\partial\lambda_1^\mu  
	\ \  \ - \ \ A_{2,31} \, 
	\partial \cdot \lambda_2 \lambda_1^\mu 
	\ +A_{3,21} \,  \partial \cdot \lambda_1 \lambda_2^\mu \\[0.8ex]
	& \qquad 	- A_{2,13} \,  \lambda_1\cdot \partial^\mu\lambda_2 \
	+A_{3,12}\, \lambda_2\cdot \partial^\mu\lambda_1 \
	\bigr]\, .
\end{split}
\ee
Since the gauge parameters appear with Grassmann odd operators, $\lambda_{12}$ computed with a symmetric vertex must be antisymmetric under  the simultaneous exchanges 
$\lambda_1 \leftrightarrow \lambda_2,  \, \bar\lambda_1 \leftrightarrow \bar\lambda_2$.  This does not happen in the above result since the local coordinates
were taken to be arbitrary. 
There are two ways to obtain the conventional multilinear function, or equivalently $\lambda_{12}^{\mu}$. One is to construct string vertices that are symmetric by design. The other is to work with arbitrary vertices and then symmetrize the value of the multilinear function. We will explore these two approaches now.

\medskip
\noindent
{\bf Symmetric vertex:} The string vertex for a three-punctured sphere 
uses
 local coordinates $w_i$, with $i = 1, 2, 3$, around the three punctures. 
 With a global coordinate $z$ describing the sphere, these local coordinates
 are defined by the maps $
z = f_i(w_i) = z_i + \alpha_i w_i + \beta_i w_i^2 + \cdots,
$
where the $z_i$ denote the locations of the punctures. 

For a symmetric vertex, we require that the local coordinates map into each other up to a phase under an $SL(2,\mathbb{C})$ transformation that permutes the three punctures. 
Without loss of generality we 
 can choose the positions 
of the punctures to be $z_1 =\beta$, $z_2= 0$, and $z_3 = -\beta$. 
The $SL(2,\mathbb{C})$ map $z \rightarrow -z$, which exchanges the first and third punctures  and fixes the puncture at $z_2 = 0$, requires that $f_2(w)$ 
 be an odd function. The $SL(2,\mathbb{C})$ map
$z \to \beta \tfrac{z - \beta}{3z + \beta}$,   
which cyclically permutes the punctures, further relates the coefficients of $f_1$ and $f_3$ to those in the expansion of $f_2$. All in all, we arrive at the local coordinates:
\be\label{3vertex}
\begin{split}
	f_1(w) & = \beta + 4a w +  \tfrac{12 a^2}{\beta} w^2 + \order{w^3}\,, \\[0.5ex]
	f_2(w) & = a w + \order{w^3}\,, \\[0.5ex]
	f_3(w) & = -\beta + 4a w - \tfrac{12 a^2}{\beta} w^2 + \order{w^3}\,. 
\end{split}
\ee
From this, we obtain:
\be
A_{i,jk} = \frac{z_{ij}}{z_{ik}} - \frac{f_j''}{2 f_j'^2} \frac{z_{ij} z_{jk}}{z_{ik}} = \tfrac{1}{2},
\ee
identically for $i,j,k \in \{1,2,3\}$, with $i\not= j\not= k$. 
To see this, take $j = 2$ for which $f''_2(0) = 0$. We then have $A_{1,23} = \tfrac{z_{12}}{z_{13}}= \tfrac{1}{2} $ and $A_{3,21} = \tfrac{z_{32}}{z_{31}}= \tfrac{1}{2}$. 
An analysis in the Appendix (see~\refb{sl2c}), shows that $A_{i,jk}$ is $SL(2,\mathbb{C})$ invariant. 
For a symmetric vertex different permutations of $i,j,k$ are related by $SL(2,\mathbb{C})$
transformations, thus giving us the same value of $A_{i,jk}$, equal to $1/2$. 
 Evaluating the primed bracket~\refb{bprimed} on the symmetric vertex then gives the algebra
\be\label{symmetric}
\begin{split}
	\lambda_{12}^\mu = \ & \ \tfrac{1}{2}(\lambda_1 \cdot \partial \lambda_2^\mu - \lambda_2 \cdot \partial \lambda_1^\mu) - \tfrac{1}{4}(\lambda_1^\mu \partial \cdot \lambda_2 - \lambda_2^\mu \partial \cdot \lambda_1) - \tfrac{1}{4}(\lambda_1 \cdot \partial^\mu \lambda_2 - \lambda_2 \cdot \partial^\mu \lambda_1) \\[1.0ex]
	+ & \ \tfrac{1}{4}(\bar{\lambda}_1 \cdot \partial \lambda_2^\mu - \bar{\lambda}_2 \cdot \partial \lambda_1^\mu) - \tfrac{1}{8}(\lambda_1^\mu \partial \cdot \bar{\lambda}_2 - \lambda_2^\mu \partial \cdot \bar{\lambda}_1),
\end{split}
\ee
which is independent of any off-shell data in the three-string vertex.  Thus, the bosonic gauge algebra
bracket is the same for all symmetric vertices. 
For $\bar\lambda^\mu_{12}$,
 simply replace $\lambda_i \leftrightarrow \bar \lambda_i$ in the right-hand side above. The first line of this result was given in~\cite{Hull:2009mi}. 
% The remaining terms for the closed SFT symmetric vertex were obtained in~\cite{zwunpub}.

\medskip
\noindent
{\bf Symmetrized vertex:} In this case, we work with non-symmetric vertices, compute correlators and finally perform a symmetrization over the punctures. The expressions to be symmetrized include terms like $A_{i,jk}$ and their antiholomorphic counterparts. Notably, we have the identity
\be
A_{i,jk} + A_{k,ji} = \frac{z_{ij}}{z_{ik}} - \frac{f_j''}{2 f_j'^2} \frac{z_{ij} z_{jk}}{z_{ik}} + \frac{z_{kj}}{z_{ki}} - \frac{f_j''}{2 f_j'^2} \frac{z_{kj} z_{ji}}{z_{ki}} = 1\,, 
\ee
 which simplifies the symmetrization of terms involving $A$ coefficients. 
 Writing the sum over the 
 six permutations as a sum over the three cyclic permutations, we have 
 \be
 \text{Sym}(A_{i,jk}) =  \tfrac{1}{6} \sum_{\pi \in S_3} A_{\pi(i),\pi(j)\pi(k)} = \tfrac{1}{6}\sum_{cyclic} ( A_{\pi(i),\pi(j)\pi(k)} + A_{\pi(k),\pi(j)\pi(i)}) = \tfrac{1}{6} \cdot 3 = \tfrac{1}{2}\,. 
 \ee
Another term appearing in the bracket and thus requiring symmetrization is
 $\bar{A}_{i,jk} \, \bar{A}_{i,kj}$.
 We denote the symmetrized result by $\bar A$:
\be 
\label{definingAbar}
\bar A = \text{Sym}( \bar{A}_{i,jk} \, \bar{A}_{i,kj}) = \tfrac{1}{6} \sum_{\pi \in S_3} \bar{A}_{\pi(i),\pi(j)\pi(k)} \, \bar{A}_{\pi(i),\pi(k)\pi(j)}\,.  
\ee
Using this, the 
conventional bracket resulting from the 
symmetrization of~\refb{bprimed} is
\be \label{symmetrized}
\begin{split}
	\lambda_{12}^\mu =\ & \ \tfrac{1}{2}(\lambda_1\cdot\partial\lambda_2^\mu - \lambda_2\cdot\partial\lambda_1^\mu) - \tfrac{1}{4}( \lambda_1^\mu\partial \cdot \lambda_2 - \lambda_2^\mu \partial \cdot \lambda_1 ) - \tfrac{1}{4}(\lambda_1\cdot \partial^\mu\lambda_2 - \lambda_2\cdot \partial^\mu\lambda_1) \\[1.0ex]
	+ &\ \tfrac{1}{4}(\bar \lambda_1\cdot\partial\lambda_2^\mu -\bar \lambda_2\cdot\partial\lambda_1^\mu ) -\tfrac{1}{2}\bar A\, ( \lambda_1^\mu\partial \cdot \bar \lambda_2 -\lambda_2 ^\mu \partial \cdot \bar \lambda_1 )\,.
\end{split}
\ee
This is the bracket for symmetrized vertices in bosonic string field theory.
For any three-string vertex the constant $\bar{A}$ 
appearing on the last term
encodes off-shell data in the gauge bracket and reduces to~$\tfrac{1}{4}$ for symmetric
vertices, as required by consistency with~\refb{symmetric}.  
For $\bar\lambda^\mu_{12}$
simply replace $\lambda_i \leftrightarrow \bar \lambda_i$ in the right-hand side above.

We demonstrate this off-shell dependence by example, as the expression for 
$\bar A$ as a sum is not particularly illuminating. Take the symmetric vertex~\refb{3vertex} and slightly deform it by the inclusion
of two constants $\beta_1, \beta_3$ 
that change the vertex when different from one:
\be
\begin{split}
	f_1(w) & = \beta + 4a\, w + \tfrac{12 a^2}{\beta} \beta_1 w^2 + \order{w^3}, \\
	f_2(w) & = a\, w + \order{w^3}, \\
	f_3(w) & = -\beta + 4a\, w - \tfrac{12 a^2}{\beta} \beta_3 w^2 + \order{w^3}.
\end{split}
\ee
For this vertex we find
\be
A_{1,23} = A_{3,21} = \tfrac{1}{2}, \quad
A_{2,13} = \tfrac{3}{2} \beta_1 - 1, \quad
A_{3,12} = 2 - \tfrac{3}{2} \beta_1, \quad
A_{1,32} = 2 - \tfrac{3}{2} \beta_3, \quad
A_{2,31} = \tfrac{3}{2} \beta_3 - 1.
\ee
From these we calculate $\bar A$, confirming that it encodes off-shell data:
$\bar{A} = \tfrac{1}{4} - \tfrac{3}{4} (\beta_1 - 1)(\beta_3 - 1)$.
Indeed, for $\beta_1 = \beta_3 = 1$, we have the general symmetric vertex and
$\bar A = \tfrac{1}{4}$.

\subsection{Diffeomorphism algebra in superstring theory}\label{difeo3ko3if}
Let us now address the computation of the Type II SFT gauge algebra, where the gauge parameters corresponding to diffeomorphisms lie in the NSNS sector. The gauge parameter string field in the NSNS sector takes the form
\begin{equation}
	\label{vertexNSgp} 
	\begin{split} 
		\Lambda_{{}_\NS}  = \int{d^Dp\over (2\pi)^D}  & \ \Bigl(-\tfrac{i}{\sqrt{2}}\lambda_\mu(p) \ 
		c\Bar{c}\, \psi^\mu\, \bar \partial\bar \xi \ e^{-\phi}e^{-2\bar \phi} \  + \tfrac{i}{\sqrt{2}}\Bar{\lambda}_\mu(p)\, c\Bar{c}\, \partial\xi \,\bar\psi^\mu \ e^{-2\phi}e^{-\bar \phi} \\
		& \hskip10pt 
		- \tfrac{1}{2} \mu(p) 
		(\partial c +\bar{\partial}\bar{c})\ c\Bar{c} \ \partial\xi\Bar{\partial}\xi\ e^{-2\phi}e^{-2\bar \phi}\ \Bigr)\ e^{ip\cdot X}\,. 
	\end{split} 
\end{equation}
Just as in bosonic string theory  the gauge parameters here are $\lambda_\mu, \bar\lambda_\mu$, and $\mu$ and the diffeomorphism and Kalb-Ramond
gauge parameters are  
$X_\mu \equiv \tfrac{1}{2}(\lambda_\mu + \bar \lambda_\mu )$ and 
$\e_\mu\equiv \tfrac{1}{2}(\lambda_\mu -\bar \lambda_\mu )$. 

The operators dual
to the states in $\Lambda_{{}_\NS}$ are:
\be
\begin{split}
	\mathcal{O}_\mu(p) &  = \sqrt{2}i\ (\partial c +\bar{\partial}\bar{c})\ c\Bar{c}\, \psi_\mu\, \bar \eta \ 
	e^{-\phi} e^{ip\cdot X}\, ,\\[1.0ex]
	\bar{\mathcal{O}}_\mu(p) & 
	= -\sqrt{2}i\ (\partial c +\bar \partial \bar c)\ c\bar c \, 
	\bar \psi_\mu\, \eta \ e^{-\bar \phi}e^{ip\cdot X}\, ,\\[1.0ex]
	\mathcal{O}(p)\  &  = -2\, c\bar c\, \eta\bar\eta \ e^{ip\cdot X} \, ,
\end{split}
\ee
satisfying: 
\be
\label{dualNSgp} 
\expval{\mathcal{O}_\mu(p), \Lambda_{{}_\NS}}  =  \lambda_\mu(-p)\, ,
\quad 
\expval{\bar{\mathcal{O}}_\mu(p), \Lambda_{{}_\NS}}  = \bar \lambda_\mu(-p)\, , \quad 
\expval{\mathcal{O}(p), \Lambda_{{}_\NS}}  = \mu(-p)\, .
\ee 
The gauge parameter in the gauge algebra~\refb{superalg} is a two component string field, with both components being identical in the NSNS sector. We now compute the field independent part of these components, which, by a slight abuse of notation, we continue to denote as $\Lambda_{12}$, as in the bosonic case. Therefore,
\be
\lambda_{12\,\mu} (-p) = \langle  {\cal O}_{\mu} (p) , [\Lambda_1 , \Lambda_2]  \rangle = \{    {\cal O}_{\mu} (p) ,  \Lambda_1 , \Lambda_2 \}  \,, 
\ee
where the evaluation of the correlator associated with the multilinear object $\{ \cdot , \cdot, \cdot \, \}$ requires the insertion of a PCO, but is otherwise defined as in the bosonic multilinear function.

We insert the PCO at an arbitrary point $z_4$, using a local coordinate $w_4$ that vanishes at this point, and then symmetrize over the PCO position under the symmetry that permutes the punctures where the string fields are inserted.

Using the primed bracket for a multilinear function computed by inserting the operators at arbitrary positions $z_1, z_2, z_3$ with local coordinates $w_1, w_2, w_3$ vanishing at these points, the correlator we must compute is
\be
\lambda'_{12\,\mu} (-p) = \{    {\cal O}_{\mu} (p) ,  \Lambda_1 , \Lambda_2 \}'  \equiv  \langle \mathcal{O}_\mu(p)(w_1)\, 
\Lambda_1(w_2)\, \Lambda_2(w_3)\,  \mathcal{X}\bar{\mathcal{X}}(w_4) \rangle \,. 
\ee
More explicitly, we have
\be
\begin{split}
	\lambda'_{12\,\mu} (-p)
	& = -\sqrt{2}i \, \langle \partial c\, c\Bar{c}\, \psi_\mu\, \bar \eta \ e^{-\phi} e^{ip\cdot X}(w_1)\, \Lambda_1(w_2)\, \Lambda_2(w_3)\,  (\partial\eta e^{2\phi}b+\partial(\eta e^{2\phi}b))e^{\bar \phi}\bar T_F(w_4)\rangle \\[1.0ex]
	& \hskip13pt- \sqrt{2}i \, \langle \bar \partial \bar c\, c\Bar{c}\, \psi_\mu\, \bar \eta \ e^{-\phi} e^{ip\cdot X}(w_1)\,  \Lambda_1(w_2),\Lambda_2(w_3)\,  e^{ \phi} T_F(\bar \partial\bar\eta e^{2\bar \phi}\bar b+\bar \partial(\bar \eta e^{2\bar \phi}\bar b))(w_4)\rangle\,, 
\end{split}
\ee
other combinations dropping out due to ghost number conservation.  
Writing out the components of the gauge parameters and keeping only terms not evaluating to zero, we get
\be
\label{diffalgtogether} 
\begin{split}
	\lambda'_{12\,\mu} (-p) =     \tfrac{i}{\sqrt{2}}\int 
	{d^D p_2\over (2\pi)^D}{d^D p_3\over (2\pi)^D}  
	 \bigl(  \lambda_{1\alpha}(p_2)\bar\lambda_{2\beta}(p_3) C_{1\mu}^{\alpha\beta}  
	+ \bar \lambda_{1\alpha}(p_2)\lambda_{2\beta}(p_3)C_{2\mu}^{\alpha\beta}  
	+\lambda_{1\alpha}(p_2)\lambda_{2\beta}(p_3) C_{3\mu}^{\alpha\beta}  \bigr) \,, 
\end{split}
\ee
with
\be
\begin{split}
	C_{1\mu}^{\alpha\beta}  = 
	& \ \langle \partial c\, c\Bar{c}\, \psi_\mu\, \bar \eta \ e^{-\phi} e^{ip\cdot X}
	\bigl| c\bar c \psi^\alpha\bar \partial \bar \xi e^{-\phi}e^{-2\bar \phi}e^{ip_2\cdot X}\bigl| c\bar c\partial  \xi\bar \psi^\beta  e^{-2\phi}e^{-\bar \phi}e^{ip_3\cdot X}
	\bigl| (\partial\eta e^{2\phi}b+\partial(\eta e^{2\phi}b))e^{\bar \phi}\bar T_F)\rangle\,,  
	\\[1.0ex]
	C_{2\mu}^{\alpha\beta}  = 
	& \ \langle \partial c\, c\Bar{c}\, \psi_\mu\, \bar\eta \ e^{-\phi} e^{ip\cdot X}
	\bigl| c\bar c\partial  \xi\bar \psi^\alpha e^{-2\phi}e^{-\bar \phi} e^{ip_2\cdot X}
	\bigl| c\bar c \psi^\beta \bar\partial\bar\xi e^{-\phi}e^{-2\bar \phi} 
	\e^{ip_3\cdot X} 
	\bigl| (\partial\eta e^{2\phi}b +\partial(\eta e^{2\phi}b))e^{\bar \phi}\bar T_F)\rangle\,, 
	\\[1.0ex]
	C_{3\mu}^{\alpha\beta}  = &  - 
	\langle \bar\partial \bar c\, c\Bar{c}\, \psi_\mu\,\bar\eta\ e^{-\phi} e^{ip\cdot X}
	\bigl| c\bar c \psi^\alpha\bar \partial \bar \xi e^{-\phi}e^{-2\bar \phi}e^{ip_2\cdot X}\bigl| c\bar c \psi^\beta\bar \partial \bar \xi e^{-\phi}e^{-2\bar \phi}
	e^{ip_3\cdot X} 
	\bigl| e^{ \phi} T_F(\bar \partial\bar\eta e^{2\bar \phi}\bar b+\bar \partial(\bar \eta e^{2\bar \phi}\bar b))\rangle\,.
\end{split}
\ee
The correlator multiplying  $\bar \lambda_1 \bar \lambda_2$ vanishes by picture number conservation.  In the correlators above, the four operators, separated by vertical bars, are inserted from left to right, at $z_1, z_2, z_3, z_4$.  
We will ignore conformal factors in~\refb{sconftrans} that only affect the algebra in terms with higher derivatives, obtaining exact, off-shell independent results for the leading terms in the algebra, that in fact have one derivative.  

A couple of useful correlators are
\be
\begin{split}
	\langle\partial c \,c\bar c(z_1)\,  c\bar c(z_2)\, c\bar c(z_3)\, b(z_4)\rangle = &   -   { z_{12}^2 \bar z_{12} \, z_{13}^2 \bar z_{13}  z_{23} \bar z_{23} \over z_{14}^2 z_{24} z_{34} } \,,  
	\\[2.0ex]
	\langle  \psi_\mu(z_1) \psi^\alpha(z_2) \psi^\beta(z_3)\psi^\nu(z_4) \rangle= &\ \frac{\delta_\mu^\alpha\eta^{\beta\nu}}{z_{12}z_{34}}+ \frac{\delta_\mu^\nu\eta^{\alpha\beta}}{z_{14}z_{23}}- \frac{\delta_\mu^\beta\eta^{\alpha\nu}}{z_{13}z_{24}} \,. 
\end{split}
\ee
The calculation of the correlators is standard but involved, requiring simplification
of terms with coordinate differences and their derivatives.  We use
combinations of coordinate differences that correspond to cross ratios, and 
momentum conservation to leave the answers in terms of $p_2$ and $p_3$.
For $C_1$ we get  
\be
C_{1\, \mu}^{\alpha\beta}  = \tfrac{1}{\sqrt{2}}(2\pi)^D\delta^{(D)}(\Sigma p) \delta_\mu^\alpha
\Bigl(\frac{\bar z_{13}\bar z_{24}}{\bar z_{14}\bar z_{23}}\, p_2^\beta  
+\frac{\bar z_{13}\bar z_{24}}{\bar z_{14}\bar z_{23}}
\frac{\bar z_{13}\bar z_{24}}{\bar z_{12}\bar z_{34}}p_3^\beta\Bigr)\,. 
\ee
Some work can be saved for $C_2$ by noticing that it is related to $C_1$
by exchanging the second and third entries in the correlator. 
This is equivalent to exchanging $z_2\leftrightarrow z_3$, $\alpha \leftrightarrow\beta$, and $p_2 \leftrightarrow p_3$, with an extra sign since the operators exchanged are odd.   We find,
\be
C_{2\, \mu}^{\alpha\beta}   = \tfrac{1}{\sqrt{2}}(2\pi)^D\delta^{(D)}(\Sigma p) \, \delta_\mu^\beta
\Bigl(\, \frac{\bar z_{12}\bar z_{34}}{\bar z_{14}\bar z_{23}}\frac{\bar z_{12}\bar z_{34}}{\bar z_{13}\bar z_{24}}\, p_2^\alpha\ 
+\frac{\bar z_{12}\bar z_{34}}{\bar z_{14}\bar z_{23}}\, p_3^\alpha  \Bigr)\,.
\ee
The third correlator is
\be
\begin{split} 
	C_{3\, \mu}^{\alpha\beta}   =  & \ \tfrac{1}{\sqrt{2}}(2\pi)^D\delta^{(D)}(\Sigma p)
	\Bigl[ \delta_\mu^\alpha \Bigl( p_2^\beta + p_3^\beta\frac{z_{13}z_{24}}{z_{12}z_{34}}\Bigr) + 
	\eta^{\alpha\beta} \Bigl(  p_{2\mu} \frac{z_{12}z_{34}}{z_{14}z_{23}} + p_{3\mu}\frac{z_{13}z_{24}}{z_{14}z_{23}} \Bigr)  
	- \delta_\mu^\beta  \Bigl( 
	p_2^\alpha \frac{z_{12}z_{34}}{z_{13}z_{24}} + p_3^\alpha\Bigr)  \Bigr]\,.
\end{split}\ee
In terms of the cross ratio $\chi$ defined by
\be
\chi \equiv   {z_{12} z_{34} \over z_{13}z_{24}} \,, \ \ \ \   1-\chi =  {z_{14} z_{23} \over z_{13}\, z_{24}}\,, 
\ee
and defining constants $c_{i\, \mu}^{\alpha\beta}$ from the relations
\be
C_{i\, \mu}^{\alpha\beta}   =  \, \tfrac{1}{\sqrt{2}}(2\pi)^D\delta^{(D)}(\Sigma p) \,  c_{i\, \mu}^{\alpha\beta} \,, 
\ee
the above results imply that
\be
\begin{split}
	c_{1\, \mu}^{\alpha\beta}  = & \ \ \delta_\mu^\alpha 
	\Bigl( \tfrac{1}{1- \bar \chi} \, p_2^\beta  +\tfrac{1}{1- \bar \chi} \, \tfrac{1}{\bar \chi} \, p_3^\beta\Bigr)\,, \\[1.0ex]
	c_{2\, \mu}^{\alpha\beta}  = & \   \delta_\mu^\beta 
	\Bigl( \tfrac{\bar \chi^2}{1- \bar \chi} \, p_2^\alpha  + \tfrac{\bar \chi}{1- \bar \chi} 
	\, p_3^\alpha\Bigr)\,,
	\\[1.0ex]
	c_{3\, \mu}^{\alpha\beta}  = & \ \delta_\mu^\alpha \bigl( p_2^\beta
	+ \tfrac{1}{\chi} \, p_3^\beta\bigr)
	+ \eta^{\alpha\beta} \bigl( \tfrac{\chi}{1-\chi} p_{2\mu}
	+ \tfrac{1}{1-\chi}\, p_{3\mu}\bigr)
	- \delta_\mu^\beta  \bigl( \chi\, p_2^\alpha
	+  p_3^\alpha \bigr)\,. 
\end{split}
\ee
We now note that
\be\lambda'_{12\, \mu} (x) = \int  {dp\over (2\pi)^D}   \lambda'_{12\, \mu} (p) e^{ipx} = 
\int  {dp\over (2\pi)^D}   \lambda'_{12\, \mu} (-p) e^{-ipx} \,. 
\ee
As a result, from \refb{diffalgtogether}, the $x$-dependent parameter is
\be
\begin{split}
	\lambda'_{12\, \mu} (x) = & \  -\tfrac{i}{{2}}\int  {dp\over (2\pi)^D} {dp_2\over (2\pi)^D}{dp_3\over (2\pi)^D}  e^{-ipx} (2\pi)^D \delta ( p + p_2 + p_3)\\
	& 
	\hskip50pt\Bigl[  \lambda_{1\alpha}(p_2)\bar \lambda_{2\beta}(p_3) 
	c_{1\, \mu}^{\alpha\beta}  
	+ \bar \lambda_{1\alpha}(p_2) \lambda_{2\beta}(p_3)c_{2\, \mu}^{\alpha\beta}  
	+ \lambda_{1\alpha}(p_2)\bar \lambda_{2\beta}(p_3) c_{3\, \mu}^{\alpha\beta}  \Bigr] \,.
\end{split}
\ee
Doing the $dp$ integral and substituting the value of the $c_i$ we have
\be
\begin{split}
	\lambda'_{12\, \mu} (x) =   & \ -\tfrac{1}{{2}}\int  {dp_2\over (2\pi)^D}{dp_3\over (2\pi)^D}  e^{ip_2x + ip_3 x} \Bigl(   \lambda_{1 \mu}(p_2)\bar \lambda_{2\beta}(p_3) 
	\Bigl( \tfrac{1}{1- \bar \chi} \, ip_2^\beta  
	+ \tfrac{1}{1- \bar \chi} \, \tfrac{1}{\bar \chi} \, ip_3^\beta\Bigr)\\
	&   + \bar\lambda_{1 \alpha }(p_2)\lambda_{2\mu}(p_3)
	\Bigl(  \tfrac{\bar\chi^2}{1- \bar \chi} \, ip_2^\alpha  
	+ \tfrac{\bar \chi }{1- \bar \chi} \, ip_3^\alpha\Bigr)  
	+ \lambda_{1\mu} (p_2) \lambda_{2\beta}(p_3)
	\Bigl(  \, ip_2^\beta  
	+ \tfrac{1}{\chi} \, ip_3^\beta \Bigr)  
	\\
	& %
	+ \lambda_{1} (p_2) \cdot \lambda_{2}(p_3)
	\Bigl(  \tfrac{ \chi }{1- \chi}\, ip_{2\mu}  
	+ \tfrac{1}{1-\chi} \, ip_{3\mu} \Bigr)  
	- \  \lambda_{1\alpha}(p_2)\lambda_{2\mu}(p_3) \bigl( \, \chi\,  ip_{2\mu}
	+ \tfrac{1}{1-\chi}\, ip_{3\, \mu}\bigr) \Bigr)\,. 
\end{split}
\ee
Using the Fourier transforms
\be
\lambda (p_i) =  \int dx  \lambda (x_i) e^{-ip_i x_i} \,,
\ee
we see that $i p = \partial$, and the integrals over $p_2, p_3$ produce position
delta functions that set all fields at~$x$.  We thus have
\be
\begin{split}
	\lambda'_{12\, \mu}  =   -{1\over 2}  \Bigl(&\  
	{1\over 1- \bar \chi}  \ \bar\lambda_2\cdot \partial \lambda_{1\mu}  + {\bar \chi
		\over 1-\bar \chi} \ \bar\lambda_1 \cdot  \partial \lambda_{2\mu}  
	+  {1\over (1-\bar \chi) \bar \chi} \  
	\lambda_{1\mu}  \partial \cdot \bar \lambda_2  
	+  {\bar \chi^2 \over 1- \bar \chi} \  \lambda_{2\mu} \partial\cdot \bar\lambda_1 
	\\ 
	&  \hskip-10pt
	+  \lambda_2 \cdot \partial  \lambda_{1\mu}   -  \lambda_1 \cdot \partial \lambda_{2\mu}  +  {1\over \chi} \  \lambda_{1\mu} - \chi \,  \partial\cdot \lambda_1  \, \lambda_{2\mu} 
	\partial\cdot \lambda_2  
	+  {\chi \over 1-\chi} \ \lambda_2 \cdot \partial_\mu \lambda_1   
	+ {1\over 1-\chi} \  \lambda_1 \cdot \partial_\mu \lambda_2  \Bigr) \,.
\end{split}
\ee
The three-string vertex in SFT is required to be symmetric under the exchange of the three punctures at $z_1$, $z_2$, and $z_3$. However, unlike the bosonic case, choosing a general symmetric vertex is not sufficient, as we also need to symmetrize over the PCO location. If we insert a PCO at the point $z_4$ with corresponding $\chi$, we need to insert at all other points which are mapped by the permutation symmetry of the three punctures $z_1,z_2,z_3$.  The permutations of three objects is a group with six elements.  Thus the three-string-vertex is a formal sum of six vertices, each with a different location of the PCO, and thus with a different value of $\chi$.  The values of the $\chi$'s are in the list $\{\chi, 1-\chi, \frac{1}{\chi}, \frac{1}{1-\chi}, 1-\frac{1}{\chi}, 1-\frac{1}{1-\chi}\}$.  Averaging over the six insertions and noting that the sum of the 
six elements in the list is equal to three, we find, 
\be
\label{supertringalgdiff}
\begin{split}
	\ \   \lambda_{12\, \mu}  \ =  & \ \ \  {\textstyle{1\over 2}} \, \bigl( \lambda_1\cdot  \partial \lambda_{2\mu}  
	- \lambda_2\cdot \partial \lambda_{1\mu} \bigr)   - {\textstyle{1\over 4}}  \bigl( \lambda_1 \cdot \partial_\mu \lambda_2 
	-   \lambda_2 \cdot \partial_\mu \lambda_1\bigr) 
	- {\textstyle{1\over 4}}  \bigl( \lambda_{1\mu} \, \partial\cdot \lambda_2  -  \lambda_{2\mu}  \, \partial\cdot \lambda_1 \bigr)      \\[0.5ex]
	& +  {\textstyle{1\over 4}}\, 
	\bigl( \bar\lambda_1 \cdot \partial \lambda_{2\mu} -
	\bar\lambda_2 \cdot \partial \lambda_{1\mu}  \bigr)  -  {\textstyle{1\over 2}}\, \bigl( \lambda_{1\mu}   \, \partial \cdot \bar\lambda_2
	- \lambda_{2\mu}  \,  \partial \cdot \bar\lambda_1 \bigr) \,. \ \ \ \phantom{\Bigl(} 
\end{split}
\ee
All terms except the last are identical to the bosonic SFT gauge algebra~\refb{symmetrized}.  
The last term is of the same form, but in bosonic string theory it comes with
a factor with off-shell data instead of $1/2$. 
For $\bar\lambda_{12\, \mu}$  simply replace in the above right-hand side $\lambda_i \leftrightarrow \bar \lambda_i$.

\medskip
\noindent
{\bf Summary of results on the bracket for gauge transformations.} 
In order to collect our results let us define a constant $\gamma$ playing the
role of $\bar A$ (as defined in~\refb{definingAbar}) for bosonic strings, and taking the value appropriate for
superstrings:   
\be
\gamma = \begin{cases} 
 \tfrac{1}{4} \,,  \ \ \ \hbox{bosonic string, symmetric vertices}\\[0.2ex]
 \bar A  \,, \ \ \ \hbox{bosonic string, symmetrized vertices}\\
 1 \,, \ \ \ \ \hbox{type II strings,  arbitrary vertices.}
\end{cases}  
\ee
We now have the algebra of transformations:
\be \label{allalgebra}
\begin{split}
	\lambda_{12}^\mu =\ & \ \tfrac{1}{2}(\lambda_1\cdot\partial\lambda_2^\mu - \lambda_2\cdot\partial\lambda_1^\mu) - \tfrac{1}{4}( \lambda_1^\mu\partial \cdot \lambda_2 - \lambda_2^\mu \partial \cdot \lambda_1 ) - \tfrac{1}{4}(\lambda_1\cdot \partial^\mu\lambda_2 - \lambda_2\cdot \partial^\mu\lambda_1) \\[1.0ex]
	+ &\ \tfrac{1}{4}(\bar \lambda_1\cdot\partial\lambda_2^\mu -\bar \lambda_2\cdot\partial\lambda_1^\mu ) -\tfrac{1}{2}
	\, \gamma \,  
( \lambda_1^\mu\partial \cdot \bar \lambda_2 -\lambda_2^\mu \partial \cdot \bar \lambda_1 )\,.
\end{split}
\ee
The result for $\bar\lambda_{12}^\mu$ is obtained by exchanging $\lambda_i \leftrightarrow \bar\lambda_i$.  
In terms of the diffeomorphism and Kalb-Ramond gauge parameters, 
$X_\mu \equiv \tfrac{1}{2}(\lambda_\mu + \bar \lambda_\mu ) $ and $\e_\mu\equiv \tfrac{1}{2}(\lambda_\mu -\bar \lambda_\mu )$ respectively, we have:
\be
\label{diffff}
\begin{split}
	X_{12\, \mu}  = \tfrac{1}{2}(\lambda_{12\, \mu} + \bar \lambda_{12\, \mu}) = &\ {\textstyle{3\over 4}} \, \bigl( X_1\cdot  \partial X_{2\mu}  
	- X_2\cdot \partial X_{1\mu} \bigr) - {\textstyle{1\over 4}}  \bigl( X_1 \cdot \partial_\mu X_2 
	- X_2 \cdot \partial_\mu X_1\bigr) \\[1.0ex]
	& - {\textstyle{1\over 4}}(1+2 \gamma ) 
	 \bigl( X_{1\mu} \, \partial\cdot X_2  - X_{2\mu} \, \partial\cdot X_1 \bigr) + {\textstyle{1\over 4}} \, \bigl( \e_1\cdot  \partial \e_{2\mu} - \e_2\cdot \partial \e_{1\mu} \bigr) \\[1.0ex]
	& - {\textstyle{1\over 4}}  \bigl( \e_1 \cdot \partial_\mu \e_2 
	- \e_2 \cdot \partial_\mu \e_1\bigr) - {\textstyle{1\over 4}} 
	(1-2\gamma)
	\bigl( \e_{1\mu} \, \partial\cdot \e_2  - \e_{2\mu} \, \partial\cdot \e_1 \bigr)\,.
\end{split}	
\ee

\be
\label{kramond}
\begin{split}
	\e_{12\, \mu}  = \tfrac{1}{2}(\lambda_{12\, \mu} - \bar \lambda_{12\, \mu}) = &\ {\textstyle{3\over 4}} \, \bigl( X_1\cdot  \partial \e_{2\mu}  
	- X_2\cdot \partial \e_{1\mu} \bigr)   - {\textstyle{1\over 4}}
	(1-2\gamma )
	\bigl( X_{1\mu} \, \partial\cdot \e_2 - X_{2\mu} \, \partial\cdot \e_1 \bigr)\\[1.0ex]
	& + {\textstyle{1\over 4}}  \bigl( \e_1 \cdot \partial X_{2\mu} 
	- \e_2 \cdot \partial X_{1\mu} \bigr)  - {\textstyle{1\over 4}}
	(1+2 \gamma) 
	\bigl( \e_{1\mu} \, \partial\cdot X_2  - \e_{2\mu} \, \partial\cdot X_1 \bigr) \\[1.0ex]
	& - {\textstyle{1\over 4}}  \bigl( X_1 \cdot \partial_\mu \e_2 
	- X_2 \cdot \partial_\mu \e_1\bigr) - {\textstyle{1\over 4}}  \bigl( \e_1 \cdot \partial_\mu X_2 
	- \e_2 \cdot \partial_\mu X_1\bigr)\,.
\end{split}	
\ee
For purely diffeomorphisms, i.e., $\e_i = 0$, we obtain the algebra:
\be
\label{supertringalgdiffXX}
\begin{split}
	X_{12\, \mu} = &\ {\textstyle{3\over 4}} \, \bigl( X_1\cdot  \partial X_{2\mu}  
	- X_2\cdot \partial X_{1\mu} \bigr) - {\textstyle{1\over 4}}  \bigl( X_1 \cdot \partial_\mu X_2 
	- X_2 \cdot \partial_\mu X_1\bigr) \\[1.0ex]
	& - {\textstyle{1\over 4}}
	(1+2\gamma )
	\bigl( X_{1\mu} \, \partial\cdot X_2  - X_{2\mu} \, \partial\cdot X_1 \bigr)\,.
\end{split}	
\ee
In the bosonic string theory 
analysis of Hull and Zwiebach~\cite{Hull:2009mi}
the original gauge transformations of the SFT, as expected,  give the algebra above with $\gamma=1/4$. These gauge transformations were simplified by field redefinitions and (field-dependent) redefinitions of the gauge parameters resulting in eqn.\,(3.22) in~\cite{Hull:2009mi}. 
The algebra was then re-computed from the simpler gauge transformations. The result
of this computation, given in~\cite{Hull:2009mi}, eqn.~(3.32), imply that:
\be
\begin{split}
	\lambda_{12} ^\mu = &\ \tfrac{1}{2} \bigl(  (\lambda_1 + \bar \lambda_1) \cdot \partial \lambda_2^\mu 
	- (\lambda_2 + \bar \lambda_2) \cdot \partial \lambda_1^\mu \bigr) \\[1.0ex]
	& - \tfrac{1}{4} \bigl( \lambda_1 \cdot \partial^\mu \lambda_2 - \lambda_2 \cdot \partial^\mu \lambda_1 \bigr) + \tfrac{1}{4} \bigl( \bar\lambda_1 \cdot \partial^\mu \bar\lambda_2 - \bar\lambda_2 \cdot \partial^\mu \bar\lambda_1 \bigr)\,.
\end{split}
\ee
For diffeomorphisms ($X = \lambda = \bar \lambda$), the terms on the second line cancel, and one recovers the expected Lie bracket:
\be
X_{12}^\mu = X_1 \cdot \partial X_2^\mu - X_2 \cdot \partial X_1^\mu = [X_1, X_2]^\mu\,.
\ee
The redefinitions for the superstring algebra are considered in~\cite{Mamade:2025jbs}.
The effective field theory gives the Lie bracket, but the SFT gives the more
complicated algebra (with $\gamma=1$).
Agreement is demonstrated by showing that these algebras are mapped to each
other by field dependent redefinitions of the
gauge parameters, as shown in section 6.1 of that paper (see eqn.(6.14)). 

\sectiono{Trivial gauge transformations and redefinitions}\label{gaderivLinf} 
		
Following standard usage (see, for example~\cite{Henneaux:1989jq}) we define gauge transformations as transformations that can be prescribed
independently at each spacetime point and leave the action invariant (up to total derivatives).   
Infinitesimal gauge transformations of the field, written as $\delta \Psi$,
leave the action invariant to first order in $\delta \Psi$.   Our focus is on
such transformations.

We have already defined in section~\ref{bossftgaualg} {\em standard gauge transformations}. 
They are of the form $\delta \Psi= Q'\Lambda$
with $\Lambda$ an arbitrary Grassmann odd string field of ghost number one.  
These transformations indeed leave the action invariant to first order in $\delta \Psi$.  

 A quantity $M$ is said to vanish on-shell, and is written as
$M\approx 0$, if $M$ vanishes when $\FF = 0$.  This notation helps us deal with
{\em trivial gauge transformations}, also defined in section~\ref{bossftgaualg}, as gauge
transformations with $\delta \Psi \approx 0$.  Trivial gauge transformations 
cannot be ignored because they are generated in string field theory by the commutator of two standard gauge transformations.   Trivial transformations form a normal
subalgebra of the algebra of gauge transformations~\cite{Henneaux:1989jq}:
the commutator of a trivial transformation with a standard one gives a trivial
transformation, and the commutator of trivial transformations must be a trivial
transformation.  

In field theory, trivial gauge transformations of field, denoted as $\phi^i$, are given in the form~\cite{Henneaux:1989jq}: 
\be\label{trivial}
\delta^{{}^T} \hskip-3pt \phi^i  =   \mu^{ij}  {\delta S \over \delta \phi^j} \,, 
\ee
with $\mu^{ij} = - \mu^{ji}$, an antisymmetric function of fields and derivatives and
the functional derivative of $S$ is the field equation.  This is clearly an invariance  of
any action.  Moreover, with some regularity conditions it is known that any gauge transformation that vanishes on-shell,  $\delta\phi^i \approx 0$ can be written in the above form, for some suitable $\mu$. 
In SFT, however, there is more structure to the trivial gauge transformations.  We have an $L_\infty$
algebra, and it is thus natural to ask what kinds of trivial gauge transformations can
be written using the equation of motion ${\cal F}$, the string field $\Psi$, and the
string products.  Below, we make some progress answering this question. 

\medskip
It should be noted that there are approaches to the gauge symmetry of
SFT in which one has a strict Lie algebra structure.  This was discussed
in~\cite{Sen:1993ic}, and further explored in~\cite{Grigoriev:1998gn},
both in the framework of Batalin-Vilkoviski quantization.
Collectively denoting fields and antifields as $\Phi^i$, transformations obtained via
the antibracket as 
\be
\delta \Phi^i = \{ \Phi^i , K\}\,,
\ee
with $K$ an odd function of the $\Phi$'s, result in a variation of the action in
the form $\delta S = \{ S, K \}$.  By the Jacobi identity of the antibracket and the classical
master equation $\{ S , S \} = 0$ we get $\delta S = 0$ if
$K$ is `trivial' in the sense that it can be written as 
\be
K = \{ S , \Lambda\}  \quad \to \quad \delta_\Lambda \Phi^i= \{ \Phi^i , \{ S, \Lambda\} \} \,. 
\ee
Here $\Lambda$ is an even function of the $\Phi$'s.  The familiar
gauge transformations of the SFT arise from such $K$'s, in fact, 
ones for which $\Lambda = \Phi^i \omega_{ij} \Lambda^k$, with 
constants $\Lambda^k$ and with $\omega$ the symplectic form, 
is a linear function of the fields/antifields.   These
transformations do not close among themselves, but once one includes
general functions $\Lambda$ one actually gets them to close.  Indeed,~\cite{Sen:1993ic} finds that
\be
[ \delta_{\Lambda_2}, \delta_{\Lambda_1} ]  = \delta_{[\Lambda_1, \Lambda_2]} %rm I think the order of 1 and 2 are opposite  \,,
\ee 
where the bracket of the functions $\Lambda_1, \Lambda_2$  is
\be
[\Lambda_1, \Lambda_2] = \tfrac{1}{2} \bigl(  \{ \Lambda_1 , \{ S, \Lambda_2\}\} 
- \{ \Lambda_2 , \{ S, \Lambda_1\}\}  \bigr) \,. 
\ee
To have a Lie algebra the bracket must also satisfy a Jacobi identity. It was shown
that the Jacobiator of the bracket does not vanish but it is a function of the form $\{ S, \chi\}$. Such a $\Lambda$-type function generates no symmetry transformation
since $\{ S, \{ S, \chi\} \} = 0$.  To make the Jacobiator strictly zero,  one must work in  the space of equivalence classes of functions
$\Lambda \sim \Lambda + \{S , \chi\}$.  It is simple to check that the bracket only
depends on the equivalence class of the gauge parameters.  Thus the bracket
defines the Lie algebra of classical gauge transformations.  It is conceivable
that one may be able to identify a Lie subalgebra that corresponds to diffeomorphisms. 

An alternative approach using coalgebras and cyclic coderivations 
for $L_\infty$ leads to the 
same conclusions.  Degree zero coderivations ${\bf v}$ that commute with the coderivation ${\bf M}$ that defines the $L_\infty$ algebra generate symmetries.
Moreover, `trivial' coderivations ${\bf v} = [{\bf M, \Lambda} ]$, defined by coderivations
${\bf \Lambda}$ include the familiar string theory gauge symmetries and close~\cite{Erler:2016rxg,Vosmera:2020jyw}.\footnote{We thank the referee of this article for sketching such argument.}

\subsection{Formalizing the identification of diffeomorphisms}\label{froma} 
We now reformulate the Ghoshal-Sen setup using $L_\infty$ algebra technology, which provides a precise characterization of gauge structures. In this language, the consistency of the EFT $\to$ SFT embedding becomes a statement about morphisms of  $L_\infty$ algebras and compatible gauge parameter mappings.   We believe that this 
approach would be applicable for theories with `exotic' versions of diffeomorphisms,
such as type IIB theory or Green-Schwarz corrected double field theory~\cite{Hohm:2014eba}. In all these
theories one could expect the embedding of standard diffeomorphisms to work,
perhaps at the cost of making other symmetries less manifest. 
It should be cautioned, however, that a complete demonstration that
standard diffeomorphisms are identifiable in SFT has not been yet carried out
even for bosonic closed strings.

\medskip
The full SFT has an $L_\infty$ structure.  The low-energy effective theory (EFT) of the massless sector, including the gravity field, is also described by an $L_\infty$ algebra. 
The requirement that these theories be related implies the existence of an $L_\infty$ morphism mapping the EFT to the SFT. We now write down the conditions in this language.
On the SFT side, we have the  string field $\Psi$, the gauge parameter $\Lambda$, and the $L_\infty$ products $b_n$ which were represented using brackets $[\cdots]$ in section~\ref{bossftgaualg}. On the EFT side, we have the collection of fields $\Phi$, the gauge parameters $\eta$, and the $L_\infty$ products $\tilde b_n$ to be represented by
$[\![\cdots ]\!]$. 
We want to find the $L_\infty$ morphism $F$ relating the theories. We have, that an arbitrary off-shell EFT field $\Phi$ is mapped to the string field $\Psi$ as follows:
\begin{equation}\label{Lmorph}
	\begin{split}
		\Psi = F(\Phi) = \sum_{n=1}^\infty\frac{1}{n!}\, f_n(\Phi^n) = f_1(\Phi) + \tfrac{1}{2}f_2(\Phi, \Phi) + \cdots\, ,
	\end{split}
\end{equation}
where the $f_n$ are symmetric multilinear maps. 
It is also useful to introduce the related map
\be
\label{F2input} 
 F (\xi, \Phi) \equiv  \sum_{n=0}^\infty  {1\over n! } f_{n+1} ( \xi, \Phi^{n} ) \,, 
\ee
with $\xi$ a field configuration of the EFT.   We then have the simple relation 
\be
\delta F ( \Phi) =  F ( \delta \Phi, \Phi) \,.  
\ee

Since $F$ is an $L_\infty$ morphism, the multilinear maps $f_n$ must relate the products in the two theories as follows~\cite{Jurco:2018sby}:
\be
\label{linftymorph}
\sum_{i+j = n} f_{i+1}\circ \tilde b_j = \sum_{i=1}^{n}\frac{1}{i!}\sum_{j_1+\cdots+j_i =n} b_j \circ (f_{j_1}\wedge\cdots\wedge f_{j_i}) \,,
\ee
where the wedge denotes the symmetrized tensor product. The first few identities are
\begin{equation}  
	\begin{split}
		f_1 \tilde b_1& = b_1 f_1\,, \\
		f_2 (\tilde b_1\otimes 1 + 1 \otimes \tilde b_1)   
		+ f_1\tilde b_2 &= b_1f_2 + b_2(f_1\wedge f_1)\,. 
	\end{split}
\end{equation}

We also look for a map from the EFT gauge parameters $\eta$ to the SFT
gauge parameters $\Lambda$.  Since all infinitesimal gauge transformations are linear in the gauge parameters, the $\Lambda$'s are linear in the $\eta$'s. 
This map, however, can also depend on the EFT fields and thus takes
the form
\begin{equation}
\Lambda = V(\eta, \Phi) = \sum_{n=0}^\infty  {1\over n!} \,  v_{n+1} ( \eta, \Phi^n) 
=  v_1(\eta) + v_2(\eta, \Phi) + \tfrac{1}{2} \,  v_3(\eta, \Phi, \Phi) + \cdots\,. 
\end{equation}
where the $v_i$ are multilinear maps, symmetric in the $\Phi$ entries.
The gauge transformations of the SFT and EFT take the following forms:
\begin{equation}
\label{30irjdf} 
	\begin{split}
	\hbox{SFT}: \quad	& \delta \Psi = \sum_{n=0}^{\infty} \frac{1}{n!}b_{n+1}(\Lambda, \Psi^n),\\
	\hbox{EFT}: \quad & \delta \Phi = \sum_{n=0}^{\infty} \frac{1}{n!}\tilde b_{n+1}(\eta, \Phi^n).
	\end{split}
\end{equation}
These gauge transformations must match 
given the mappings $F$ and $V$ 
introduced above. 
From the SFT transformation we have 
\begin{equation}\label{gtcon}
		\delta \Psi   
		= \sum_{n=0}^{\infty} \frac{1}{n!}b_{n+1}(V(\eta, \Phi), (F(\Phi))^n)\,. 
		\end{equation}
On the other hand, given that $\Psi = F (\Phi)$, we also have that 
\begin{equation}\label{gtchain}
		\delta \Psi  = \delta F(\Phi) =   F ( \delta_\eta \Phi, \Phi) \,, 
\end{equation}
with $\delta_\eta \Phi$ given in~\refb{30irjdf}. 
The consistency condition is the agreement of the final right-hand sides in
the equations~\refb{gtcon} and~\refb{gtchain} above:  
\be
\label{gtconscond} 
 \sum_{n, k=0}^{\infty} 
\frac{1}{n! k!}\, b_{n+1}\bigl(v_{k+1}(\eta, \Phi^{k}), (F(\Phi))^n\bigr) = 
\sum_{n,k=0}^{\infty}\frac{1}{n!k!}\, f_{n+1} \bigl(\tilde b_{k+1}(\eta, \Phi^k\bigr), 
\Phi^n)  \,. 
\ee
This gives an infinite set of conditions indexed by the action on $\eta\otimes \Phi^{k}$,
with $ k=0,1,2,\cdots$. The first two conditions are
\begin{equation}
	\begin{split}
		b_1 \circ v_1& = f_1\circ \tilde b_1 \hskip73pt \hbox{on} \quad (\eta)\, ,\\
		b_1\circ v_2 + b_2\circ(v_1\otimes f_1) & = f_1 \circ \tilde b_2 + f_2(\tilde b_1\otimes 1) \quad \hbox{on} \quad (\eta \otimes \Phi)\, .
	\end{split}
\end{equation}
Note that the last term in the second line does not have the symmetric
structure of the $L_\infty$ morphism condition~\refb{linftymorph}, where $\tilde b_1$  
acts on both entries symmetrically.  
 
The program 
succeeds if we can find  maps $f_i$ and $v_i$
that satisfy~\refb{gtconscond}.  
These are the maps that define $F$ and $V$, respectively.  Of
course, we also have to satisfy the $L_\infty$ condition~\refb{linftymorph}.  
Satisfying~\refb{gtconscond} means that not only we have embedded off-shell 
configurations of the EFT into the SFT, but we also have consistently
identified SFT gauge
parameters $\Lambda = V(\eta, \Phi)$ that generate the transformations associated with EFT gauge transformations with parameter $\eta$.   
More conceptually, we have the following commutative diagram:
\be
\label{commdiag}
 \begin{tikzcd}
	\Phi  \arrow{r}{F} 
	 \arrow[swap]{d}{\rm EFT\, gt} 
	& \Psi \arrow{d}{\rm SFT\,  gt} \\%
	\Phi + \delta_\eta \Phi \arrow{r}{F}& \Psi + \delta_{V(\eta, \Phi)} \Psi 
\end{tikzcd}
\ee
 In the bottom arrow, $F$ maps keeping only the linear part of $\delta_\eta \Phi$.
 This commutative diagram can be encoded in the relation
 \be  \delta_{V(\eta,\Phi)} F (\Phi)\ = \ F (\delta_\eta \Phi, \Phi )  \,. 
 \ee
   Structurally, we have the following relation of operations acting on $\Phi$
 \be
 \label{conceptcond}
   \delta_{V(\eta, \Phi)} \,  F \ = \   F \,  (\delta_\eta \Phi,  \, \boldsymbol{\cdot} \,  )   \,. 
 \ee  
 One recognizes the left-hand and right-hand sides of~\refb{gtconscond} correspond to the
 left-hand and right-hand sides of the above equation, respectively.

It was found in~\cite{Ghoshal:1991pu}, however,  that even to first order in fields, the conditions can only be satisfied if the SFT gauge transformations
include the trivial gauge transformations: 
\begin{equation}
	\begin{split}
		\delta_\Lambda \Psi = \sum_{n=0}^{\infty} \frac{1}{n!}b_{n+1}(\Lambda, \Psi^n) + K(\Lambda, \mathcal{F}(\Psi), \Psi)\, ,
	\end{split}
\end{equation}
where $ K(\Lambda, \mathcal{F}(\Psi), \Psi)$  
is linear in $\Lambda$ 
and linear in the equation of motion $ \mathcal{F}(\Psi)$.  
This transformation, when written in terms of component fields, 
is of the form of~\refb{trivial}.  This extra term changes the explicit form of the condition~\refb{gtconscond}, but~\refb{conceptcond} is unchanged; only
the expression for $\delta_{V(\eta, \Phi)}$ changes.   

\medskip 
It is also 
of interest to consider the SFT and EFT algebras of gauge transformations and their
consistency. 
Here are some preliminary 
comments on this point. The SFT algebra, as calculated
earlier,~is 
\begin{equation}\label{SFTga}
	\begin{split}
		[\delta_{\Lambda_1}, \delta_{\Lambda_2}] \Psi & = \delta_{[\Lambda_1,\Lambda_2]'}\Psi + [\Lambda_1, \Lambda_2, \mathcal{F}]' + {\rm trivial}\, ,
	\end{split}
\end{equation}
with additional trivial transformations that may be required.
Given that the EFT is an $L_\infty$ theory we have
\be
\label{lbefdiffs}
[\delta_{\eta_1}, \delta_{\eta_2} ] \Phi = 
\delta_{[\![ \eta_1, \eta_2 ]\!]'} \Phi  + [\![ \eta_1, \eta_2 , \tilde {\mathcal{F}}]\!]'\, .
\ee
where $[\![\cdots ]\!]$ denote the $L_\infty$ products of the EFT 
and $\tilde {\cal F}$ is the EFT equation of motion.  
To relate the above transformations we consider~\refb{conceptcond} 
\be
\delta_\Lambda \Psi = \delta_{V(\eta,\Phi)}  F (\Phi) = 
F ( \delta_\eta\Phi , \Phi) \,.   
\ee
Then we find that 
\begin{equation} 
\label{nfeiuc}
	\begin{split}
		[\delta_{\Lambda_1 }, \delta_{\Lambda_ 2}] \Psi  = [\delta_{V(\eta_1, \Phi) }, \delta_{V( \eta_2,\Phi)}] F(\Phi) 
		& \ = \ 
\delta_{V(\eta_1, \Phi) }\Bigl(\delta_{V( \eta_2,\Phi)} F(\Phi)\Bigr) - (1\leftrightarrow2) \qquad \\[1.0ex]
& = \delta_{V(\eta_1, \Phi) }
 F( \delta_{\eta_2} \Phi, \Phi )- (1\leftrightarrow2)\\[1.0ex]
&  =  F \bigl(\delta_{[\![ \eta_1, \eta_2 ]\!]'} \Phi\,, \Phi\bigr) 
+ F\bigl( [\![ \eta_1, \eta_2 , \tilde {\mathcal{F}}]\!]' \,, \Phi \bigr) \\[1.0ex]
		&  = \delta_{[\![ \eta_1, \eta_2 ]\!]'} F(\Phi) + F( [\![ \eta_1, \eta_2 , \tilde {\mathcal{F}}]\!]', \Phi ) \\[1.0ex]
&= \delta_{V( [\![ \eta_1, \eta_2 ]\!]', \Phi)} \Psi 
+ F( [\![ \eta_1, \eta_2 , \tilde {\mathcal{F}}]\!]' \,, \Phi)  \,. 
	\end{split}
\end{equation}
Comparing with the SFT gauge algebra~\refb{SFTga}, we find, as would have
been expected, 
\begin{equation}
\label{conc03}
 [\Lambda_1,\Lambda_2]'  = V( [\![ \eta_1, \eta_2 ]\!]', \Phi) = 
  [V(\eta_1, \Phi),V(\eta_2,\Phi)]'  \,. 
\end{equation}
In the case of diffeomorphisms, equation~\refb{lbefdiffs} becomes
\be
[\delta_{\eta_1}, \delta_{\eta_2} ] \Phi = 
\delta_{[\eta_1, \eta_2]_L} \Phi  \,,   
\ee
with $[\eta_1,\eta_2]_L$  the Lie bracket of vector fields. This implies that~\refb{conc03}
becomes 
\begin{equation}
\label{conc04}
  V( [ \eta_1, \eta_2 ]_L, \Phi) =  
  [V(\eta_1, \Phi),V(\eta_2,\Phi)]'  \,. 
\end{equation}
For diffeomorphisms the map $V$ should satisfy the above condition.  
Moreover, for diffeomorphisms the final right-hand side of~\refb{nfeiuc}
does not include the last term, proportional to $\tilde {\cal F}$, thus
the trivial transformations we added to the SFT gauge transformations
should be such that the trivial part of the gauge algebra vanishes. Clearly,
the complete classification of  trivial transformations would be needed.

\subsection{Trivial gauge transformations} 

A key motivation to study trivial gauge transformations is that the original
gauge transformations of the SFT can be redefined to include trivial gauge transformations.  This inclusion will modify the gauge algebra and in particular
the original trivial transformations, with the possibility of eliminating them in
the case of diffeomorphisms.  As we consider possible trivial transformations
in SFT we also need to be able to compute their algebra.

To better understand trivial gauge transformations we note that there are some
particularly simple ones, transformations $\delta \Phi = Q'\Lambda$ with $ \Lambda \approx 0$.  Namely,  the gauge parameter vanishes on shell. 
There is not much to say about these; $\Lambda$ is built with products that
involve ${\cal F}$ at least linearly.   We will focus on trivial transformations
in which $\delta \Psi \approx 0$ but cannot be written as standard gauge
transformations with a parameter that vanishes on-shell by using the identities
of the $L_\infty$ algebra.   In the following, we will simply call these trivial gauge
transformations. 

We examined two forms of trivial transformations in~\refb{two-kind-of-trivial},
which we copy here including right-hand sides that follow from the $L_\infty$ identities~\refb{jyndlcsbtt}:
\be
\label{two-kind-of-trivialx} 
\begin{split}
\delta_{\chi}^{{}^T} \Psi \ \equiv \ & \   [ \chi\,  {\cal F} ]' \ = \ 
-Q' (Q'\chi) \,, 
\\[1.0ex]
 \delta^{{}^T}_{\Lambda_1, \Lambda_2  } \Psi \ \equiv \ & \   [ \Lambda_1 \Lambda_2 {\cal F} ]'  =  -  \bigl(   Q' [ \Lambda_1 \Lambda_2]'  +  [Q'\Lambda_1  \Lambda_2]'  + (-1)^{\Lambda_1} \, [\Lambda_1 Q'\Lambda_2 ]'  \bigr) \,. 
\end{split}
\ee
Clearly $\delta \Psi \approx 0$ and these are invariances of the action $S$, 
by the usual argument 
that uses $\delta S = \langle \delta \Psi , {\cal F}\rangle$, the graded commutativity
of~\refb{stillgradedcomm}, and the Grassmann odd property of ${\cal F}$.  
On the final right-hand sides we see that the $L_\infty$ identities do not
allow them to be rewritten as standard gauge transformations with gauge
parameters vanishing on-shell. 
  It is of interest to find the full set of such trivial gauge transformations.

\medskip
\noindent
{\bf A generating set of trivial gauge transformations?}  
 We present here a discussion that gives a large set of 
generating trivial transformations, but make no claim that this is the
full set.  In order to have a trivial gauge transformation $\delta \Psi$ must
be written with $L_\infty$ products that involve the string field $\Psi$, the field
equation ${\cal F}$, at least linearly, and some gauge parameters.  What is
nontrivial to guarantee is that such $\delta \Psi$ leaves the action invariant. 

The transformations in~\refb{two-kind-of-trivialx} 
are the first two members of an infinite class of trivial gauge transformations
\be
\label{firstclas-tr}
\delta_n \Psi =  [ \rho_1 \cdots  \rho_n \, \FF ]'  \,, \ \ \  \hbox{with} \ \ \ 
n = 1 ,2 , \cdots \, .
\ee
Here the $\rho$'s denote string fields of definite Grassmanality and ghost number,
satisfying conditions that make $\delta \Psi$ of ghost number two and Grassmann even.
Clearly $\delta \Psi \approx 0$ and the usual argument shows it leaves the action
invariant.   In the above one could have used the un-primed, original, brackets and still have a trivial symmetry.  We use primed brackets because those are the
ones that arise in the original gauge algebra computation.

While one could imagine these are all the trivial gauge transformation, by computing the commutator of a $\delta_1$ with a $\delta_2$, which must be trivial, one finds
additional trivial transformations.  Consider the transformations
\be
\label{30ierjfdl}
\delta\Psi =  \ [ \rho_1 \rho_2 [ \rho_3 {\cal F} ]' ]'\,  \, \pm \, \,  [ \rho_3 [ \rho_1 \rho_2 {\cal F} ]']'  \,, 
\ee
with a sign, indicated by $\pm$, that is fixed in terms of the Grassmanalities of the
$\rho_i$ but need not be specified to understand how this is a trivial gauge transformation.   To see this, just consider the first term of the above transformation
when appearing in $\langle \delta \Psi, {\cal F} \rangle$:
\be
\langle  [ \rho_1 \rho_2 \underbrace{ [ \rho_3 {\cal F} ]}\,  ]'\, , \, \underbrace{{\cal F}} \rangle = \pm \langle \, \underbrace{ [ \rho_1 \rho_2 {\cal F} \,  ]'} \, , \,  [ \rho_3 \underbrace{ {\cal F}} ]' \rangle
= \, \pm \langle \, {\cal F} \,, \,  [ \rho_3 [ \rho_1 \rho_2 {\cal F} ]']'  \rangle\,,
\ee
where at each step we exchanged the under-braced terms as allowed by graded
commutativity.  Since the final result is the ${\cal F}$ contracted with the
second term in~\refb{30ierjfdl}, this shows that a cancellation resulting in a 
vanishing variation of the action will arise for some choice of the sign in $\delta \Psi$.
This shows $\delta \Psi$ is a trivial symmetry, and a look at $L_\infty$ relations
quickly convinces one that it cannot be written as a standard gauge transformation with a gauge parameter that vanishes on-shell. 

It is not hard to generalize this argument to show that the following is
a trivial gauge transformation:
\be
\label{mostgenetypeI}
\delta \Psi =  [ x_1 \cdots [ x_2 \cdots [  \cdots   [ \cdots [ x_k \cdots {\cal F}]']' \hskip-2pt\cdot\hskip-1pt\cdot \hskip-1pt ]' \,  \pm\,  [ x_k \cdots [ x_{k-1} \cdots [  \cdots   [ \cdots [ x_1 \cdots {\cal F}]']' 
\hskip-2pt\cdot\hskip-1pt\cdot \hskip-1pt ]'\, . 
\ee
Here the dots following the $x$'s are extra arbitrary inputs that appear until the
next product arises, nesting consecutively 
until ${\cal F}$ is
found at the end.  Consecutive moves based on graded commutativity show this
is an invariance of the action.  The previous example in~\refb{30ierjfdl} was a case of this general form with $k=2$.  
One can view~\refb{firstclas-tr} as a kind of $k=1$ transformation, where
the second term is not needed.  The transformation~\refb{mostgenetypeI} is the most general trivial transformation we consider here.  
It remains to be seen if additional 
trivial transformations are needed for a complete classification.

\medskip
\noindent
{\bf Extended gauge transformations}:  They are the sum of a standard
gauge transformation with parameter $\Lambda$  and a 
trivial gauge transformation as in~\refb{two-kind-of-trivialx} 
with parameter $\chi$ of ghost-number zero:
\be
\label{extgatitio}
\delta^{\,{}^E}_{\Lambda, \chi } \,\ \equiv \ \delta_\Lambda \,  \, + \, 
\delta_{\chi}^{{}^T} \,,
\ee
resulting in 
\be
\label{jveorser}
\delta^{\,{}^E}_{\Lambda, \chi } \,\Psi \ = \ Q' \Lambda \, + \, 
[\chi\, {\cal F}\,  ]' \,. 
\ee
It is useful to note that such extended gauge transformation
is simply a standard gauge transformation with a more complicated parameter.
For this first note that  
\be
 \delta_{\chi}^{{}^T} \Psi \ = \   [{\cal F} \,  \chi  ]'  \ = \ -Q' (Q' \chi)   \ = \ 
 - \delta_{ Q'\chi}  \Psi , 
\ee
showing that a trivial transformation with parameter $\chi$
is in fact an ordinary gauge transformation with parameter $-Q'\chi$, and implying that the extended transformations in~\refb{extgatitio} are  
\be
\delta^{{}^{\rm E}}_{\Lambda, \chi}  = \delta_{\wt\Lambda} \,, \ \ \ 
\wt\Lambda \equiv  \ \Lambda - Q'\chi\, , \ \ \ \   
\delta^{{}^{\rm E}}_{\Lambda, \chi}  \Psi  =  
  Q' (\Lambda - Q'\chi) = [\Lambda - [\chi]']'\,. 
\ee
\medskip
\noindent
Now for the algebra of such extended transformations we have
\be
\label{exttransgauguesdflk}
\begin{split}
	[\delta^{{}^{\rm E}}_{\Lambda_2, \chi_2},   \delta^{{}^{\rm E}}_{\Lambda_1, \chi_1} ]\Psi 
	&\  =\  \delta_{\wt \Lambda_2}   [ \Lambda_1 -[ \chi_1]' ]'   -(1\leftrightarrow 2)\\[1.0ex]
	& \ = \  [ (\Lambda_1 -[ \chi_1]') \delta_{\wt \Lambda_2} \Psi  ]' - [ [ \chi_1, \delta_{\wt \Lambda_2} \Psi]' ]'  -(1\leftrightarrow 2)\\[1.0ex]
	& \ = \  [\,  \wt\Lambda_1  Q'\wt \Lambda_2 \,  ]' - [ [ \chi_1 Q' \wt \Lambda_2]' ]'  -(1\leftrightarrow 2)\,. 
\end{split}
\ee
Using the $L_\infty$ identities we find
\be
\begin{split}
	[\delta^{{}^{\rm E}}_{\Lambda_2, \chi_2},   \delta^{{}^{\rm E}}_{\Lambda_1, \chi_1} ]\Psi 
	&\  =\  \ Q'[\, \wt \Lambda_1 , \wt \Lambda_2]' 
	+ [\wt \Lambda_1 , \wt \Lambda_2, \FF]'
	- Q'[\chi_1, Q'\wt \Lambda_2]' + Q'[\chi_2, Q'\wt \Lambda_1]'\\[1.0ex]
	& \ = \ Q' \bigl( [\, \wt \Lambda_1 , \wt \Lambda_2]' 
	-[\chi_1, Q'\wt \Lambda_2]' + [\chi_2, Q'\wt \Lambda_1]'\bigr) 
	+ [\wt \Lambda_1 , \wt \Lambda_2, \FF]'\,.
\end{split}
\ee
With the identification $\delta^{{}^{\rm E}}_{\Lambda_i, \chi_i}  = \delta_{\wt\Lambda_i}$
$(i=1,2$),  the gauge algebra has become
\be
[ \delta_{\wt\Lambda_2}, \delta_{\wt \Lambda_1}] = \delta_{ \wt\Lambda_{12}}
+ \delta^{{}^{T}}_{\wt\Lambda_1, \wt\Lambda_2}  \,,  
\ee
where 
\be 
\wt \Lambda_{12} = [\wt \Lambda_1 , \wt \Lambda_2]'- [\chi_1, Q'\wt\Lambda_2]' + [\chi_2, Q'\wt\Lambda_1]'.
\ee
The new bracket, defined by $\wt\Lambda_{12}$ is the old bracket  $[\wt \Lambda_1 , \wt \Lambda_2]'$   modified 
by the addition of $\chi_i$ dependent contributions.   This means that
{\em standard} gauge transformations with field dependent 
gauge parameters $\wt \Lambda_1$ and $\wt \Lambda_2$ commute to give
a standard gauge transformation with parameter $\wt\Lambda_{12}$ which
depends on $\wt \Lambda_1,\Lambda_2, \chi_1$, and $\chi_2$.   The gauge 
algebra `bracket' has been changed, but there are still trivial gauge symmetries 
in the commutator.   At the field independent level, the algebra bracket
has been modified~as 
\be 
\wt \Lambda_{12} = [\wt \Lambda_1 , \wt \Lambda_2]- [\chi_1, Q\wt\Lambda_2] + [\chi_2, Q\wt\Lambda_1]  + {\cal O}(\Psi).
\ee
This illustrates how the bracket in the 
gauge algebra has off-shell information:  it can be changed by
alteration of the gauge parameters.   

\medskip
\noindent
{\bf Redefining gauge parameters.} We now consider a variation in the above calculations in order to directly
consider redefinitions of the gauge parameters.  
In this case we introduce a new parameter $\wt \Lambda$,  linear on the
original parameter $\Lambda$, but including all powers of the string field.
Schematically, we have 
\be
\wt \Lambda =   \Lambda  -  \sum_{n=1}^\infty c_n \Lambda \Psi^n \,.  
\ee
If we try to implement this with the string products one could write
\be
\label{naiveattempt}
\wt \Lambda =   \Lambda  -  [ (B_1\Lambda) \Psi]  -   [ (B_2\Lambda) \Psi\Psi ] + \cdots\,,
\ee
where $B_1, B_2, \cdots$ are operators of ghost number minus one, needed for the
consistency of the expansion, as every term in the series must have ghost number one,
just like $\Lambda$ does.  One can easily work out the new algebra of gauge
parameters here and, to leading field-independent order,  one finds that 
$[\delta_{\wt\Lambda_2} , \delta_{\wt\Lambda_1} ] = \delta_{[\wt\Lambda_1, \wt\Lambda_2]_*} + \cdots$ where the new bracket $[\  ,\,  ]_*$ is the following
modification  
\be
[ \wt \Lambda_1 , \wt \Lambda_2 ]_* = [ \wt \Lambda_1 , \wt \Lambda_2 ] -
[ B\wt \Lambda_1 , Q\wt\Lambda_2 ] + [ B\wt \Lambda_2 , Q\wt\Lambda_1 ]  + {\cal O} (\Psi)  \,.
\ee
It is possible to place the $B$ operators at places different than the one
shown in~\refb{naiveattempt}, such as $B [ \Lambda \Psi]$ or as 
$B [ \Lambda \Psi]$.  In those cases, the new bracket takes different but related
forms.   This is what we do in practice
at the component level to recover the Lie bracket as the bracket of
diffeomorphisms (\cite{Hull:2009mi} for bosonic strings and \cite{Mamade:2025jbs} for
type II strings).

In order to work more systematically, 
 we can identify a set of ghost number zero  parameters 
$\xi_i (\Lambda)$,
all linear in $\Lambda$, that
could be used to redefine the parameters as in
\be
\label{naiveattempt2}
\wt \Lambda =   \Lambda  -  [ \xi_1(\Lambda) \Psi]  -   [ \xi_2(\Lambda) \Psi\Psi ] + \cdots\,.
\ee
For the sake of simplicity and to get insight from analytic expressions,
 let us consider the case where there is just one parameter
$\xi (\Lambda)$, linear in $\Lambda$ and field independent,
 such that the redefinition takes the form  
\be
\label{naiveattempt3} 
\wt \Lambda =   \Lambda  - [ \xi\Psi]  -  \tfrac{1}{2}    [ \xi \Psi^2 ] + \cdots\,.
\ee
We declare that the terms in dots take the precise form that allows us to use $Q'$ to write :
\be
\wt \Lambda =  \Lambda + Q\xi  \ - \bigl( Q\xi  + [\xi  \Psi ]  + \tfrac{1}{2} [ \xi , \Psi^2] + \cdots \bigr)  =   \Lambda + Q\xi   - Q' \xi   \,.
\ee
The associated transformation is
\be
\delta_{\wt \Lambda} \Psi =  Q' ( \Lambda + Q\xi )  +  [ \xi {\cal F} ]'  \,,   
\ee
so that comparing with~\refb{jveorser} we see that this is in fact an extended
gauge transformation  with parameters 
\be
\delta_{\wt \Lambda}  = \delta^{{}^E}_{ \Lambda + Q\xi\, , \,  \xi } \,.
\ee
The computation of the gauge algebra of $\delta_{\wt \Lambda}$'s  is almost identical to that starting in~\refb{exttransgauguesdflk}.  We write $\xi_i = \xi (\Lambda_i)$ and find   
\be
\label{exttransgauguesdflka}
\begin{split}
	[\delta_{\wt\Lambda_2},   \delta_{\wt\Lambda_2} ]\Psi 
	&\  =\  \delta_{\wt \Lambda_2}   [ \Lambda_1+ Q\xi_1 -[ \xi_1]' ]'   -(1\leftrightarrow 2)\,,  \\[1.0ex]  
	& \ = \  [\,  \wt\Lambda_1  Q'\wt \Lambda_2 \,  ]' - [ [ \xi_1 Q' \wt \Lambda_2]' ]'  -(1\leftrightarrow 2)\,,\\[1.0ex]   
	& \ = \ Q' \bigl( [\, \wt \Lambda_1 , \wt \Lambda_2]' 
	-[\xi_1 Q'\wt \Lambda_2]' + [\xi_2 Q'\wt \Lambda_1]'\bigr) 
	+ [\wt \Lambda_1  \wt \Lambda_2, \FF]'\,.
\end{split}
\ee
The bracket is changed, as expected:
\be
\wt \Lambda_{12} =  [\, \wt \Lambda_1 , \wt \Lambda_2]' 
	-[\xi_1 Q'\wt \Lambda_2]' + [\xi_2 Q'\wt \Lambda_1]'\,.
\ee
and the algebra can be written as 
\be
[\delta_{\wt\Lambda_2},   \delta_{\wt\Lambda_2} ]  =  \delta_{\wt\Lambda_{12}} 
+ \delta^{{}^T} _{ \wt \Lambda_1 , \wt \Lambda_2}\,.  
\ee
The trivial transformation is not changed, at first sight.  But the 
familiar ambiguity remains, in which we can modify the gauge parameter
and the trivial transformation:
\be
\delta_{\wt\Lambda_{12}}  =  \delta_{\wt\Lambda_{12} - Q'\rho_{12}}  + \delta^{{}^T}_{\rho_{12} }\,. 
\ee
For some suitable $\rho_{12}$, this gives an avenue to both modify the bracket by the addition of $(-Q'\rho_{12})$, 
and the trivial transformations of the algebra by the addition of $[{\cal F } \rho_{12}]$
to the transformation of the field.  It is conceivable that this can be used to 
simplify or perhaps even eliminate the trivial transformations.

A more radical approach could be to modify the gauge transformations
by the explicit addition of a trivial transformation that cannot be traded by
a standard gauge transformation, as in 
\be
\wt\delta_{\Lambda } \Psi =  Q' \Lambda  +  [\Lambda, \Lambda_0, {\cal F}] +\cdots
\ee
with $\Lambda_0$ some fixed string field.   The algebra of such transformations
involves a rather complicated collection of trivial symmetries, for which our
first steps in classification could be of some practical~use.

		\bigskip
		\noindent
		{\bf Acknowledgements}.  We would like to thank Atakan Hilmi Firat and Ashoke Sen for helpful discussions.  We are grateful to the referee, whose questions
		on the identification of standard diffeomorphisms in SFT led to
		the exposition in section~\ref{froma} and the comments 
		immediately preceding 
		that section. The work of RAM is supported by the MIT Dean
		of Science Fellowship and MIT Department of Physics. The work of B.Z was supported
		by the U.S. Department of Energy, Office of Science, Office of High Energy Physics of
		U.S. Department of Energy under grant Contract Number DE-SC0012567.(High Energy
		Theory research). %rm added funding source

		\appendix

		\sectiono{Conformal transformations}
		
		In this appendix, we collect the conformal transformations of states that are used throughout the document. For primary operators, the transformation under a holomorphic map $ f $ is given by:
		\begin{equation}
			f\circ V(w,\bar w) = (f'(w))^h(\bar f'(\bar w))^{\bar h}V(f(w),\bar f(\bar w)).
		\end{equation}
		We now consider two non-primary vertex operators: $ \partial X^\mu e^{ip\cdot X}$ and $ (\partial^2 c)\, c $. 
		
		For $ \partial X^\mu e^{ip\cdot X} $, we compute the conformal transformation via point-splitting, using the transformations of the primary components:
		\begin{equation}
			\begin{split}
				f\circ \partial X^\mu(w) &= f'(w)\partial X^\mu(f(w)), \\
				f\circ e^{ip\cdot X}(w) &= (f'(w))^{p^2/4}(\bar f'(\bar w))^{p^2/4}e^{ip\cdot X}(f(w)).
			\end{split}
		\end{equation}
		From the definition of normal ordering:
		\begin{equation}
			\partial X^\mu e^{ip\cdot X}(w) = \lim_{z\rightarrow w} \left( \partial X^\mu(z) e^{ip\cdot X}(w) + \frac{1}{2}\frac{ip^\mu}{z-w}e^{ip\cdot X}(w) \right).
		\end{equation}
		Applying the conformal map to both sides, we find:
		\begin{equation}
			\begin{split}
				f\circ :\partial X^\mu e^{ip\cdot X}(w): &= \lim_{z\rightarrow w} \left( f\circ \partial X^\mu(z) :e^{ip\cdot X}(w): + \frac{1}{2}\frac{ip^\mu}{z-w} f\circ :e^{ip\cdot X}(w): \right) \\
				&= \lim_{z\rightarrow w} (f'(w))^{p^2/4}(\bar f'(\bar w))^{p^2/4} \bigg( f'(z): \partial X^\mu(f(z)) e^{ip\cdot X}(f(w)): \\
				&\hspace{30pt} - f'(z)\frac{1}{2}\frac{ip^\mu}{f(z)-f(w)} :e^{ip\cdot X}(f(w)): + \frac{1}{2}\frac{ip^\mu}{z-w} :e^{ip\cdot X}(f(w)): \bigg).
			\end{split}
		\end{equation}
		Evaluating the limit of the last two terms:
		\begin{equation}
			\begin{split}
				\frac{f'(z)}{f(z)-f(w)} - \frac{1}{z-w} &= \frac{1}{z-w} \left[ \frac{f'(w)+(z-w)f''(w) + \cdots}{f'(w)+\tfrac{1}{2}(z-w)f''(w) + \cdots} - 1 \right] 
				 \frac{f''(w)}{2 f'(w)} + \order{z-w}.
			\end{split}
		\end{equation}
		Now, taking the limit:
		\begin{equation}
			f\circ \partial X^\mu e^{ip\cdot X}(w) = (f'(w))^{p^2/4+1}(\bar f'(\bar w))^{p^2/4}\left( \partial X^\mu e^{ip\cdot X}(f(w)) - \frac{ip^\mu}{4} \frac{f''(w)}{f'(w)^2} e^{ip\cdot X}(f(w)) \right).
		\end{equation}
		Starting from the transformation of the primary field $ c(w) $ one derives the following
		\begin{equation}
			\begin{split}
				f\circ (\partial c\, c)(w) &= (f'(w))^{-1} \partial c\, c(f(w)), \\
				f\circ (\partial^2 c\, \partial c\, c)(w) &= \partial^2 c\, \partial c\, c(f(w)), \\
				f\circ (\partial^2 c\, c)(w) &= \partial^2 c\, c(f(w)) - \frac{f''(w)}{f'(w)^2} \partial c\, c(f(w)).
			\end{split}
		\end{equation}
		Collecting all these, the relevant conformal transformations for bosonic states are:
		\be\label{conftranf}
		\begin{split}
			f \circ {\bar \partial}^2\bar c\, \partial c\, c\, \bar c\, \partial X^\mu  e^{ip\cdot X}(w) &= \abs{f'(w)}^{p^2/2} \Bigl( {\bar \partial}^2\bar c\, \partial c\, c\, \bar c + \frac{\bar f''}{\bar f'^2} \partial c\, \bar \partial \bar c\, c\, \bar c \Bigr) \Bigl( \partial X^\mu - \frac{ip^\mu}{4} \frac{f''}{f'^2} \Bigr) e^{ip\cdot X}(f(w)), \\
			f \circ {\bar \partial}^2\bar c\, \bar \partial \bar c\, c\, \bar c\, \partial X^\mu  e^{ip\cdot X}(w) &= \abs{f'(w)}^{p^2/2} \left( {\bar \partial}^2\bar c\, \bar \partial \bar c\, c\, \bar c \right) \Bigl( \partial X^\mu - \frac{ip^\mu}{4} \frac{f''}{f'^2} \Bigr) e^{ip\cdot X}(f(w)), \\
			f\circ (c\, \partial X^\mu e^{ip\cdot X})(w) &= \abs{f'(w)}^{p^2/2}  c\, \Bigl( \partial X^\mu - \frac{ip^\mu}{4} \frac{f''}{f'^2} \Bigr) e^{ip\cdot X}(f(w)) .
		\end{split}
		\ee
		In the superstring sector, all relevant operators are primary, and their transformations are thus straightforward:
		\be\label{sconftrans}
		\begin{split}
			f \circ \left( \partial c\, c\, \bar c\, \psi_\mu\, \bar \eta\, e^{-\phi} e^{ip\cdot X} \right)(w) &= \abs{f'(w)}^{p^2/2} \left( \partial c\, c\, \bar c\, \psi_\mu\, \bar \eta\, e^{-\phi} e^{ip\cdot X}(f(w)) \right), \\
			f \circ \left( c\, \bar c\, \partial \xi\, \bar \psi^\alpha\, e^{-2\phi} e^{-\bar \phi} e^{ip\cdot X} \right)(w) &= \abs{f'(w)}^{p^2/2} \left( c\, \bar c\, \partial \xi\, \bar \psi^\alpha\, e^{-2\phi} e^{-\bar \phi} e^{ip\cdot X}(f(w)) \right).
		\end{split}
		\ee

		\medskip  
		\noindent
		We now show that $A_{i,jk}$, defined in~\refb{aijk},   
		is unchanged by $SL(2,\mathbb{C})$ transfomations.  Consider
		\be
		z' = \frac{A z + B}{C z + D}, \qquad AD - BC = 1.
		\ee
		Local coordinates around the puncture $z_i$ are expanded as:
		\be
		z = f_i(w) = z_i + b_i w + c_i w^2 + \cdots,
		\ee
		with \( z_i = f_i(0) \), \( b_i = f_i'(0) \), and \( c_i = \tfrac{1}{2} f_i''(0) \).
Then under the $SL(2,\mathbb{C}) $ 
transformation:
		\be 
		\begin{split}
		z' = & \ \frac{A f_i (w) + B}{C f_i(w)  + D} \\
		= & \  \frac{A z_i + B}{C z_i + D} + \frac{b_i}{(C z_i + D)^2} w + \left( \frac{c_i}{(C z_i + D)^2} - \frac{C b_i^2}{(C z_i + D)^3} \right) w^2 + \cdots \equiv  g_i (w) \,,  
		\end{split}\ee
		where $g_i(w)$ denotes the local coordinate around $z_i'$. 
		We read the expansion coefficients:
		\be  
			z_i' = g_i(0) = \frac{A z_i + B}{C z_i + D}, \qquad 
			g_i'(0) =  {b_i\over (Cz_i+ D)^2} \,, \ \ \  
			g_i''(0) = 
			2 \left( \frac{c_i}{(C z_i + D)^2} - \frac{C b_i^2}{(C z_i + D)^3} \right).
		\ee
		Hence,
		\be
		\frac{g_i''(0)}{2 g_i'(0)^2} = \frac{f_i''(0)}{2 f_i'(0)^2} (C z_i + D)^2 - C (C z_i + D).
		\ee
		For coordinate differences,
		\be
		z'_{ij} = z'_i - z'_j = \frac{z_{ij}}{(C z_i + D)(C z_j + D)}.
		\ee
		Under the above $ SL(2,\mathbb{C}) $ map $z_i \rightarrow z'_i = g_i(0) $, we find that, as claimed, 
		\be\label{sl2c}
		\begin{split}
			A'_{i,jk}  = 
			\frac{z'_{ij}}{z'_{ik}} - \frac{g_j''}{2 g_j'^2} \frac{z'_{ij} z'_{jk}}{z'_{ik}} &= \frac{C z_k + D}{C z_j + D} \frac{z_{ij}}{z_{ik}} - \frac{f_j''}{2 f_j'^2} \frac{z_{ij} z_{jk}}{z_{ik}} + \frac{C}{C z_j + D} \frac{z_{ij} z_{jk}}{z_{ik}} \\
			&= \frac{z_{ij}}{z_{ik}} - \frac{f_j''}{2 f_j'^2} \frac{z_{ij} z_{jk}}{z_{ik}}= 
			A_{i, jk} \, .
		\end{split}
		\ee
		The same holds for the antiholomorphic sector. 
Since the $A'$s are invariant, so is the quantity $\bar A$, thus establishing 
the required $SL(2,\mathbb{C}) $ invariance of the computation of the off-shell
gauge algebra.

	\end{document}